\documentclass[useAMS,usenatbib]{mn2e}
\usepackage{graphicx}
\usepackage{rotating,times,pictex,graphicx,latexsym}
\usepackage{color}
\usepackage{longtable,amsmath}
\usepackage{lscape}

\title{A systematic survey for infrared star clusters with $|b|\,<\,20^\circ$
using 2MASS} 

\author[Froebrich, Scholz \& Raftery]{D.~Froebrich$^{1,4}$\thanks{E-mail:
df@star.kent.ac.uk}, A.~Scholz$^2$ and C.L.~Raftery$^3$ \\ $^1$ Dublin Institute for
Advanced Studies, 5 Merrion Square, Dublin 2, Ireland \\ $^2$ University of
Toronto, 60 St. George Street, Toronto, Canada \\ $^3$ Trinity College Dublin,
College Green Dublin 2 Ireland \\ $^4$ Centre for Astrophysics and
Space Science, University of Kent, Canterbury, CT2 7NQ, UK}

\begin{document}

\date{Received sooner; accepted later}
\pagerange{\pageref{firstpage}--\pageref{lastpage}} \pubyear{2006}
\maketitle

\label{firstpage}

\begin{abstract}

We used star density maps obtained from 2MASS to obtain a sample of star
clusters in the entire Galactic Plane with $|b|<20^\circ$. A total of 1788 star
cluster candidates are identified in this survey. Among those are 681
previously known open clusters and 86 globular clusters. A statistical analysis
indicates that our sample of 1021 new cluster candidates has a contamination of
about 50\,\%. Star cluster parameters are obtained by fitting a King profile to
the star density. These parameters are used to statistically identify probable
new globular cluster candidates in our sample. A detailed investigation of the
projected distribution of star clusters in the Galaxy demonstrates that they
show a clear tendency to cluster on spatial scales in the order of 12-25\,pc, a
typical size for molecular clouds.

\end{abstract}


\section{Introduction}

Star clusters are the building blocks of galaxies. By investigating the basic
properties of clusters, e.g. morphology, mass function, and spatial
distribution, we can trace the long-term evolution of stars in the Milky Way.
The pre-requisite for such studies are large-scale surveys providing the means
to detect large homogeneous samples of open (OpCl) and globular (GlCl)
clusters.

So far, about 1100 OpCls are known in the Galaxy, and their properties have
been compiled in large OpCl databases (see e.g. Mermilliod \& Paunzen
\cite{2003A&A...410..511M}, Lynga \cite{1995yCat.7092....0L}, Dias et al.
\cite{2002A&A...389..871D}). Most of these clusters, however, have been
detected in the framework of optical surveys, mainly based on photographic
plates. Since interstellar and cloud extinction are a strong factor at
wavelengths shorter than 1\,$\mu$m, optical cluster catalogues may be highly
incomplete. Near infrared (NIR) surveys are probably more appropriate to
provide a more complete census of clusters, since they are able to probe more
distant regions in the Galaxy due to much lesser sensitivity to interstellar
extinction, and to detect the youngest clusters, which are still embedded in
molecular clouds. Recent studies have shown that these embedded infrared open
clusters might outnumber optical visible clusters by an order of magnitude (see
Lada \& Lada \cite{2003ARA&A..41...57L} for a recent review). Infrared OpCls
are thus an important complement to the classical cluster catalogues.  

Similarly, the known sample of GlCls is very likely incomplete because it is
mostly  based on visual inspection of photographic plates at optical
wavelengths. Of the $\sim 150$  GlCls known in our Galaxy (e.g., Harris
\cite{1996AJ....112.1487H}), only very few have been discovered using infrared
data (e.g., Hurt et al. \cite{2000AJ....120.1876H},  Kobulnicky et al.
\cite{2005AJ....129..239K}). Similar to OpCls, however, many GlCls remain
undiscovered due to obscuration from dust close to the Galactic Plane. Based
on  the spatial distribution  of the known GlCls, Ivanov et al.
\cite{2005A&A...442..195I} estimate a lower limit  of $10 \pm 3$ for the number
of unknown GlCls near the Galactic Plane ($|Z| \le 0.5$\,kpc)  and within
3\,kpc from the Galactic Centre. Moreover, the sample of off-plane GlCls is
likely to be incomplete as well, as indicated by the recent serendipitous
discoveries of the  two off-plane clusters GC\,Whiting1 (Carraro
\cite{2005ApJ...621L..61C}) and ESO\,280-SC06 (Ortolani et al.
\cite{2000A&A...361L..57O}). This is relevant, because a complete GlCl sample 
is of fundamental importance as a probe for the variety of processes that may
have contributed  to the formation of the  Galaxy: rapid protogalactic
collapse, accretion, cannibalism, galaxy collisions, star bursts  (e.g., van
den Bergh \cite{1993ApJ...411..178V}; West et al. \cite{2004Natur.427...31W}).
The diversity of these processes is obviously reflected by the diversity of
GlCls, implying that a complete census of the GlCls is crucial for the
understanding of the formation of the Galaxy.

The NIR all sky survey 2MASS (Skrutskie et al. \cite{2006AJ....131.1163S})
provides an excellent database to identify and analyse large samples of
infrared  clusters. Recent work by Dutra et al. \cite{2003A&A...400..533D} and
Bica et al. \cite{2003A&A...404..223B} proved that 2MASS indeed contains
hundreds of previously unidentified infrared clusters. A systematic search for
2MASS clusters in the entire Galactic Plane, as well as a comprehensive
analysis of their properties inferred from NIR photometry alone, has not yet
been carried out. In a previous paper, we presented a relative extinction map
of the Galactic Plane ($|b| < 20^\circ$) derived from star counts in the 2MASS
catalogue (Froebrich et al. \cite{2005A&A...432L..67F}). The original aim of
this project was to search for dark globules. It required obtaining a star
density map of the entire region. Such a map is additionally well-suited to
detect open and globular clusters. In this paper, we report about a systematic
search and analysis of Galactic star clusters based on 2MASS star counts.

The paper is structured as follows. In Sect.\,\ref{selection} we introduce our
cluster selection method and discuss the reliability of our cluster candidates.
A description and discussion of the spatial distribution of the clusters is put
forward in Sect.\,\ref{overall}. Finally we describe our method of
statistically classifying the newly detected cluster candidates to search for
the best new Galactic GlCl candidates (Sect.\,\ref{chapter4}).

\section{Cluster selection}

\subsection{Method and results}
\label{selection}

We selected all stars from the 2MASS point source catalogue with Galactic
Latitude $|b| < 20^\circ$. To ensure high photometric accuracy only objects
with a quality flag better than 'C' (i.e. only stars with a photometric
accuracy better than five times the signal-to-noise ratio) are used. The area
with $|b| < 20^\circ$ was then divided in 288 regions of 10$^\circ$ length in
Galactic Longitude and 5$^\circ$ height in Galactic Latitude. In each of these
fields, the completeness limit was determined as the peak in the histogram of
the brightness distribution of all stars in each of the three filter bands
JHKs. The star density at this brightness was then measured in each of the
bands. Independent of the completeness limit the star density was determined by
counting the number of stars in 3.5'x3.5' sized boxes every 20". 

The star density maps were searched automatically for local density
enhancements using the SExtractor software (Bertin \& Arnouts
\cite{1996A&AS..117..393B}). As cluster candidates we selected all objects with
a local star density above the 4\,$\sigma$ level of the background and an
extent of more than eleven square arcminutes. Only objects detected either in J
and H, or in H and K where selected. Objects detected only in one of the three
bands, or in J and K only, where discarded. The redundant information in
different filters allow us a reliable object identification and safely exclude
spurious detections. This automated procedure resulted in the detection of 961
cluster candidates. 

A random inspection of the star density maps, however, revealed that the
automatic detection missed quite a number of cluster-like objects slightly
fainter than the 4\,$\sigma$ detection limit and detected several objects that
are obviously not star clusters (spikes from bright stars, dark cloud edges).
This is caused by density gradients and variable S/N in the
10$^\circ$x5$^\circ$ fields due to Galactic structure and dust clouds. Hence,
an additional manual selection was performed by visually inspecting the JHKs
star density maps.

In a first step, all obvious non-cluster objects where rejected from the
automatically obtained catalogue. These rejected objects are mostly edges of
dark clouds in regions with steep star density gradients close to the Galactic
Plane. We rejected 179 objects from the automatically detected sample. Then, to
obtain our final source sample, we manually selected all remaining objects
possessing the same visual appearance in the star density maps as known star
clusters. As for the automated detection procedure, the cluster had to be
visible in the H-band and at least in one of the other two bands. This resulted
in a total of 1788 cluster candidates. 782 (44\,\%) of those were detected by
the automated procedure. The ratio of the number of clusters detected manually
and automatically is more or less constant along the Galactic Plane. Hence, the
manual selection procedure introduces no additional selection effects that
change systematically along the Galactic Plane. 

We searched the SIMBAD database\footnote{This cross identification was
performed in May 2006.} for known clusters within three times the core radius
(see Sect. \ref{propertyanalysis}) of each candidate. All entries with
classification as GlCl?, Cl*, GlCl, OpCl, *inCl, Gl?, Cl*, GlC, OpC, As* or *iC
where selected. We identified 681 known OpCls and 86 known GlCls among the
cluster candidates in our survey field. This includes e.g. the GlCl Glimpse-C01
which was recently discovered with Spitzer data by Kobulnicky et al.
\cite{2005AJ....129..239K}. 

\begin{figure*}
\fbox{\includegraphics[height=4cm,bb=0  0 610 600]{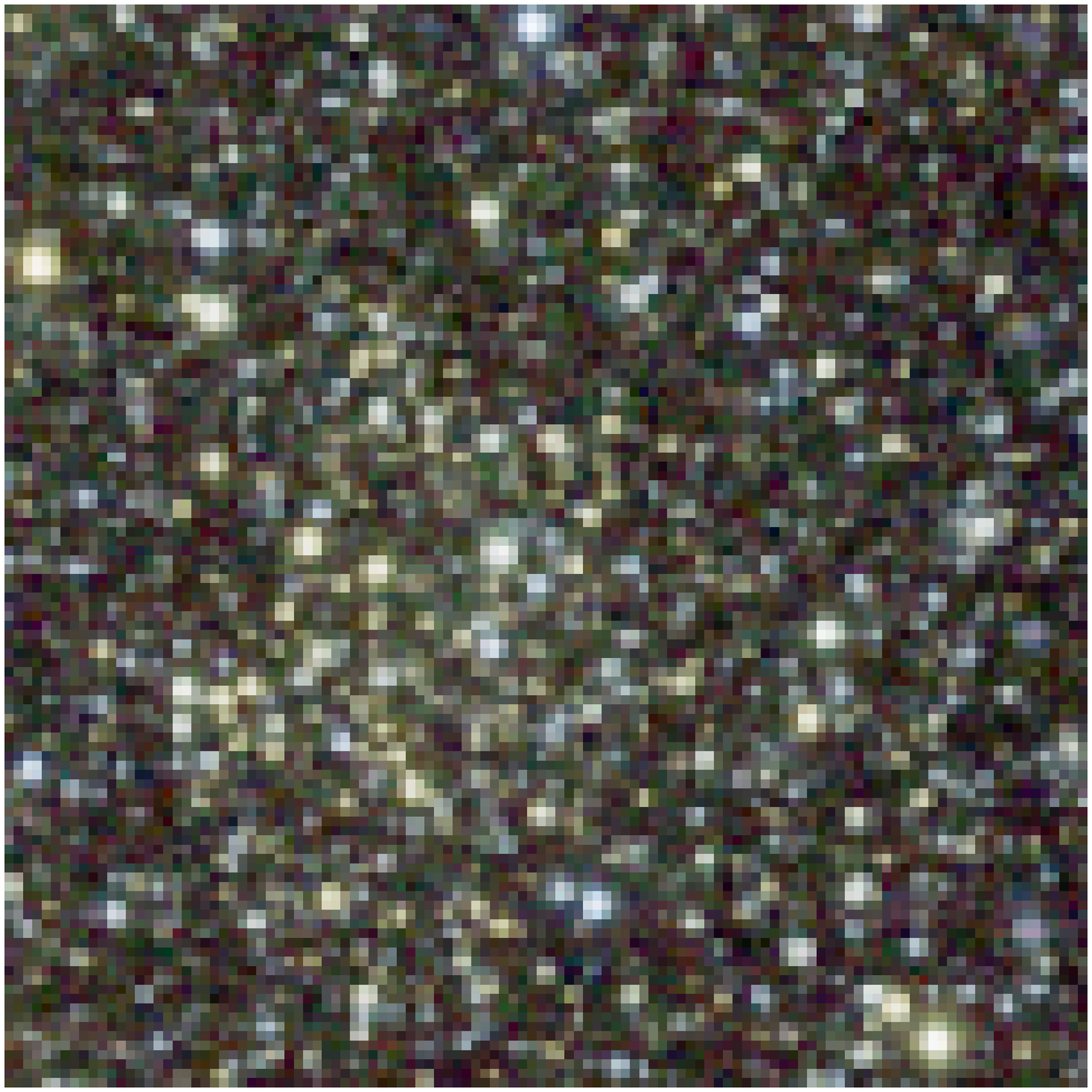}}
\fbox{\includegraphics[height=4cm,bb=0  0 720 720]{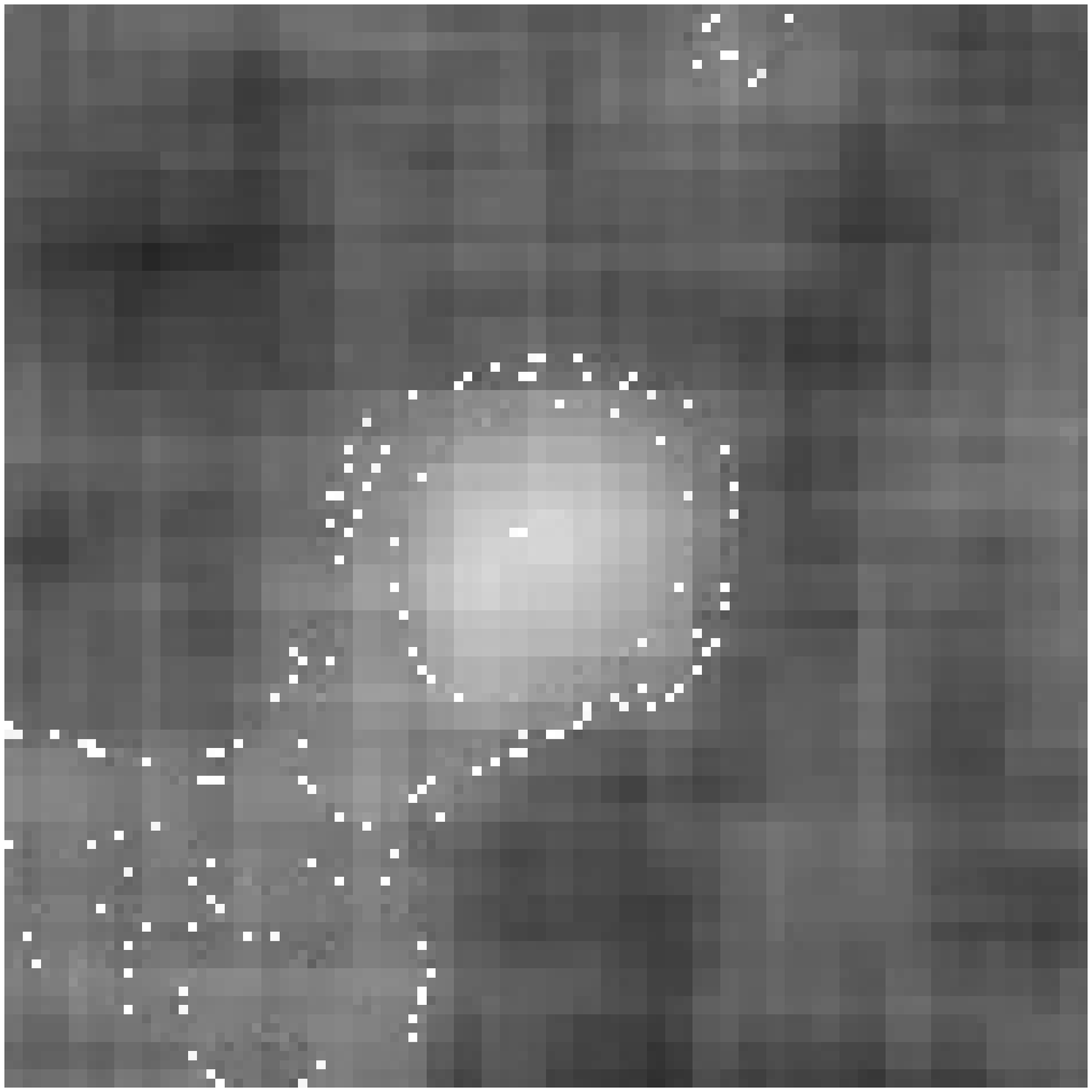}}
\fbox{\includegraphics[height=4cm,bb=0  0 610 700]{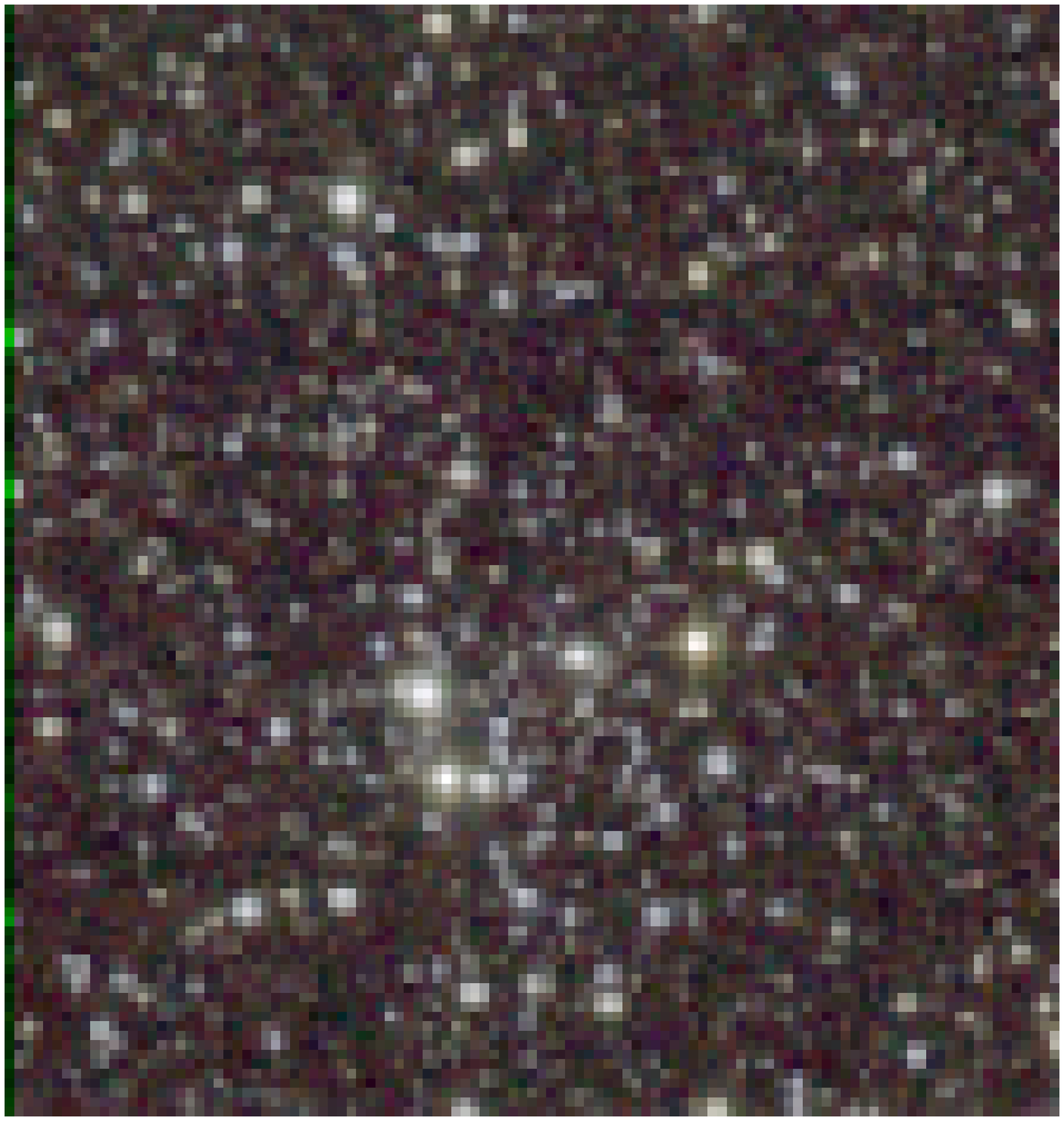}}
\fbox{\includegraphics[height=4cm,bb=0  0 720 720]{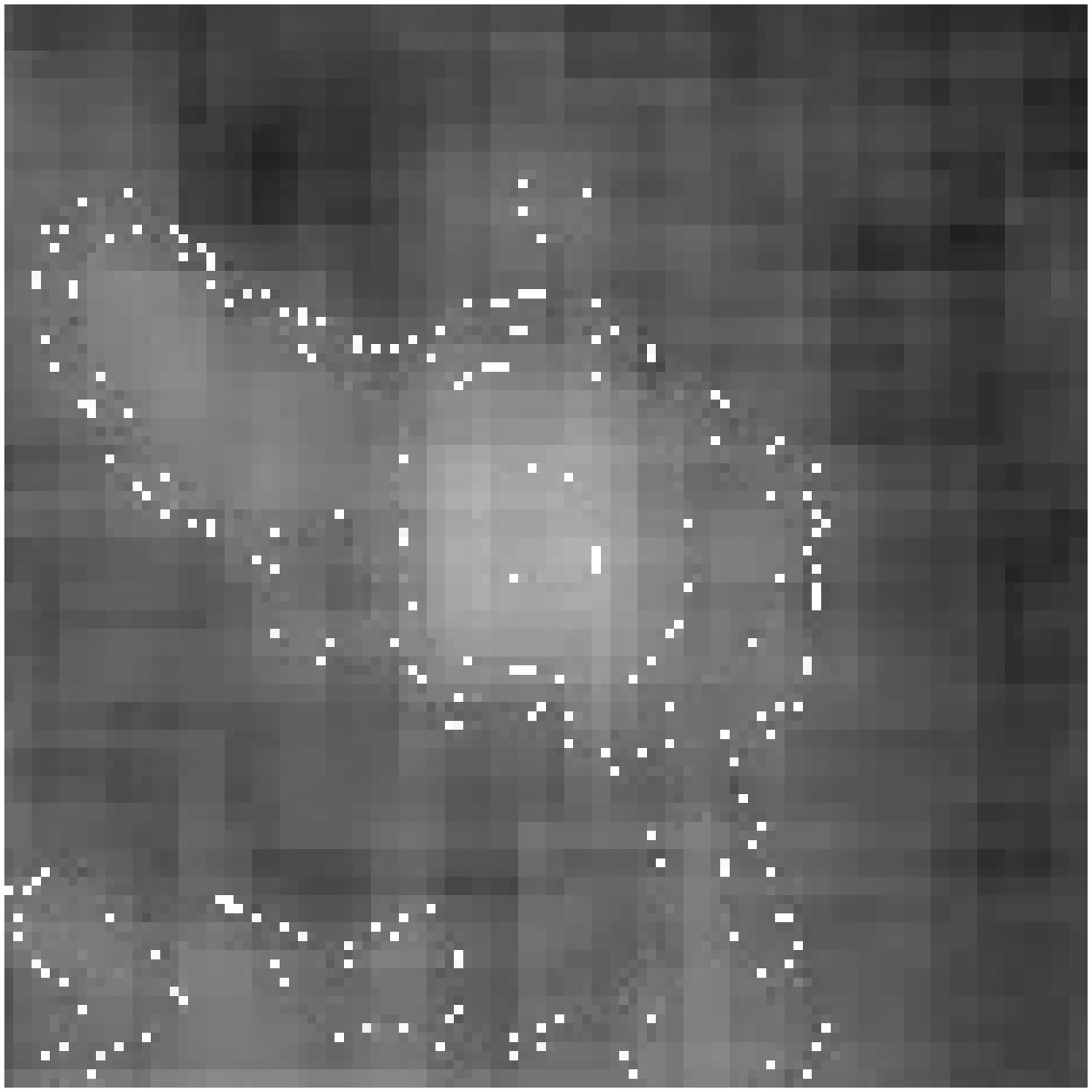}} \\
\fbox{\includegraphics[height=4cm,bb=0  0 610 700]{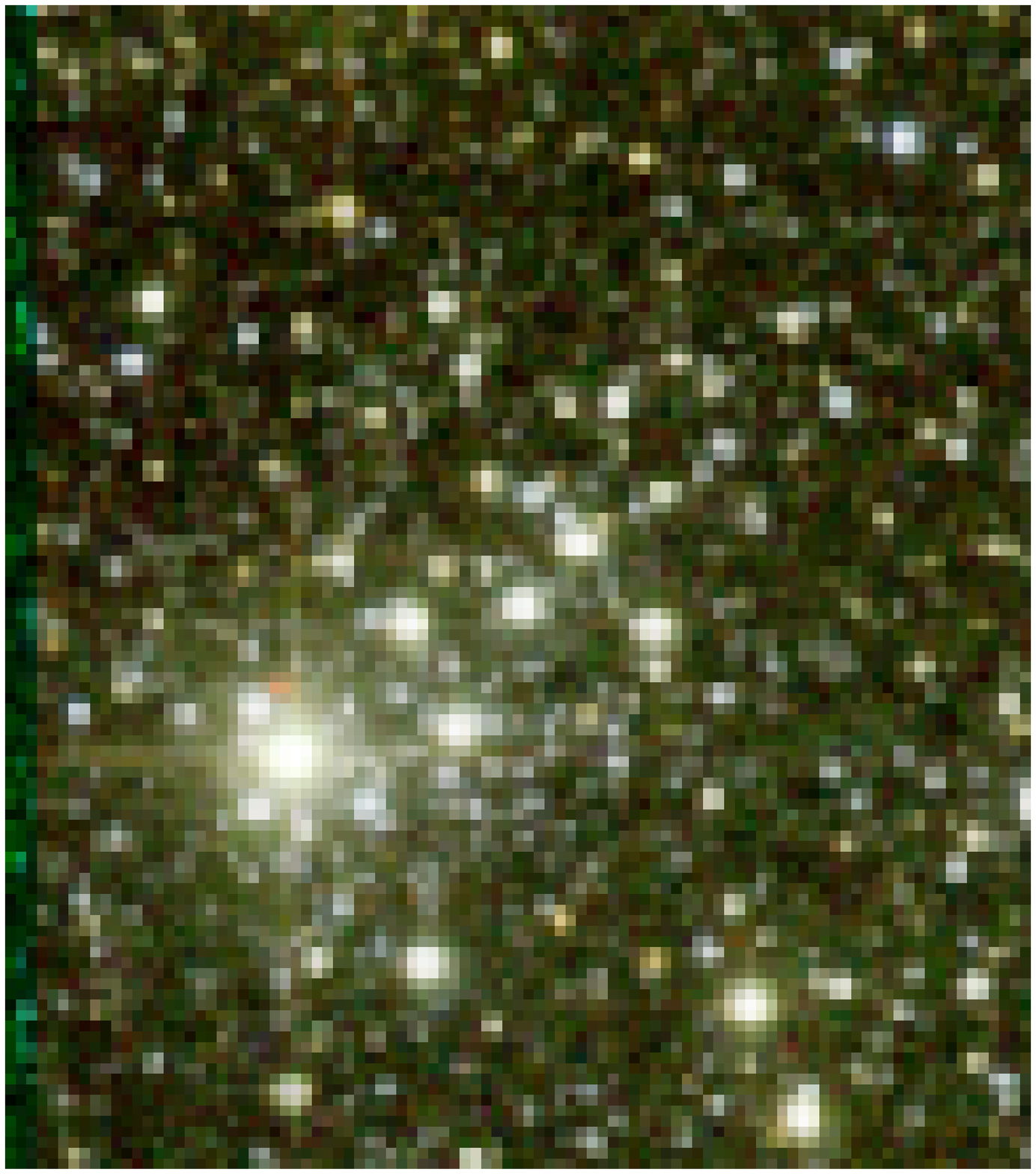}}
\fbox{\includegraphics[height=4cm,bb=0  0 720 720]{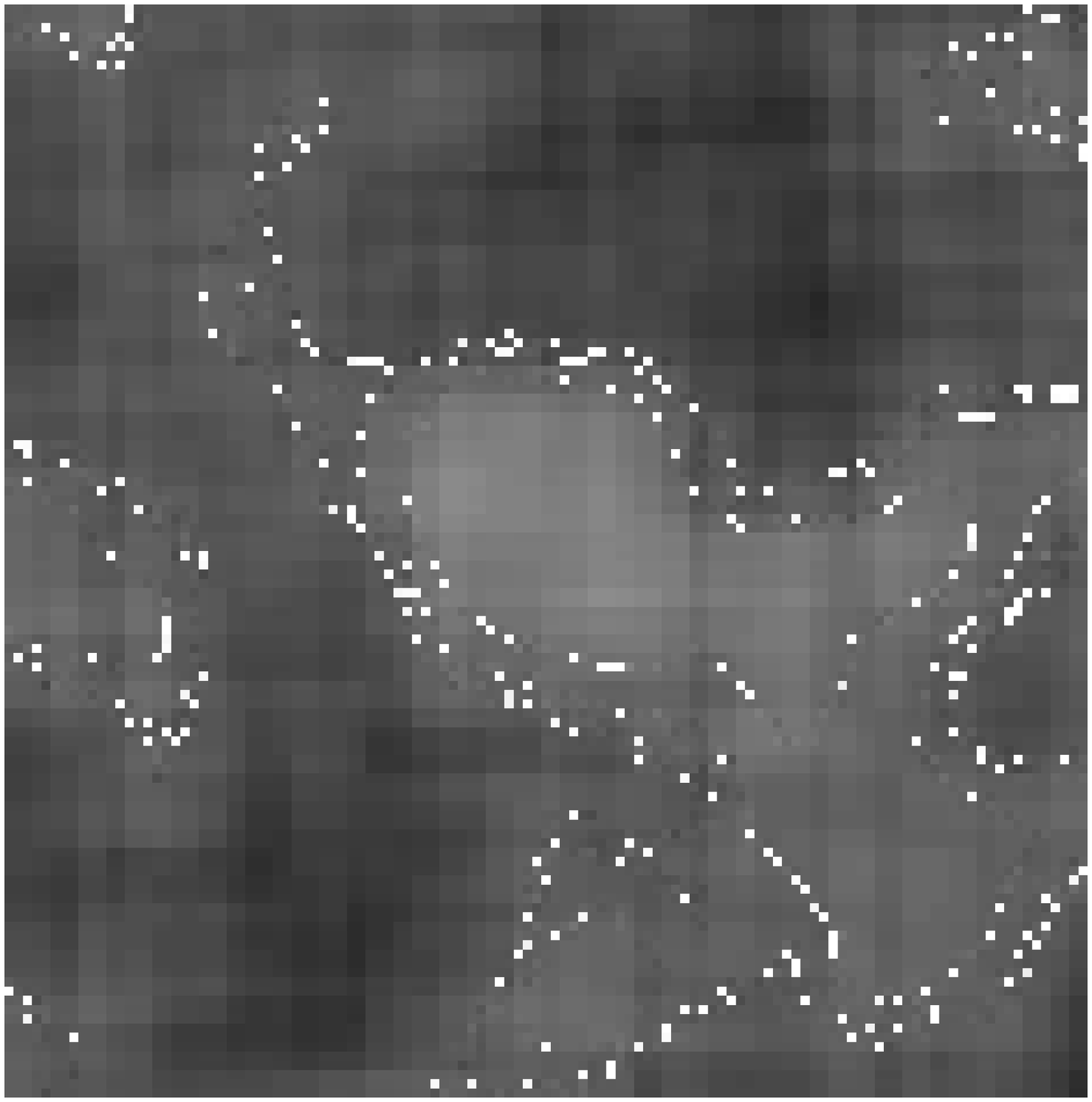}}
\fbox{\includegraphics[height=4cm,bb=0  0 610 700]{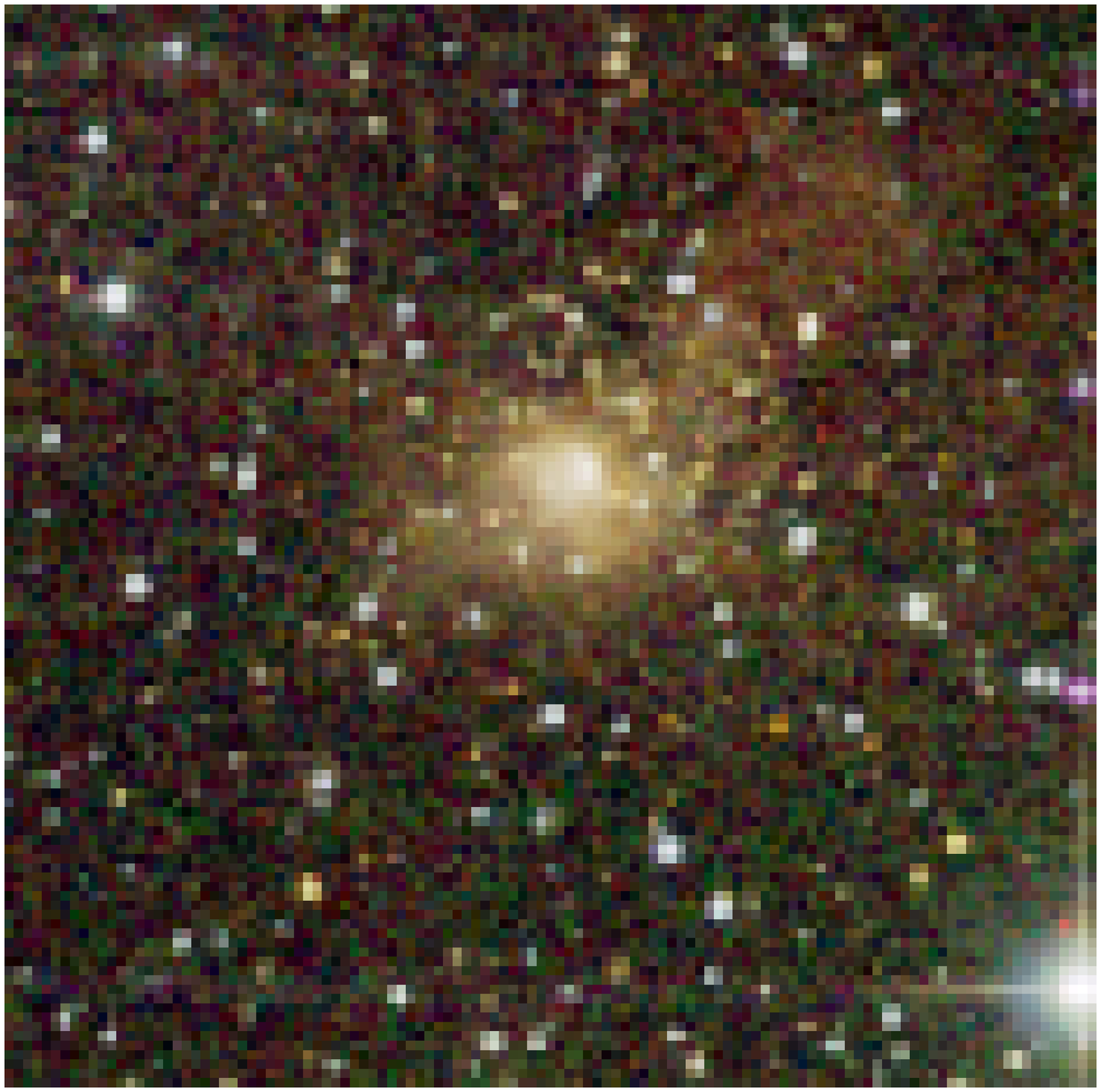}}
\fbox{\includegraphics[height=4cm,bb=0  0 720 720]{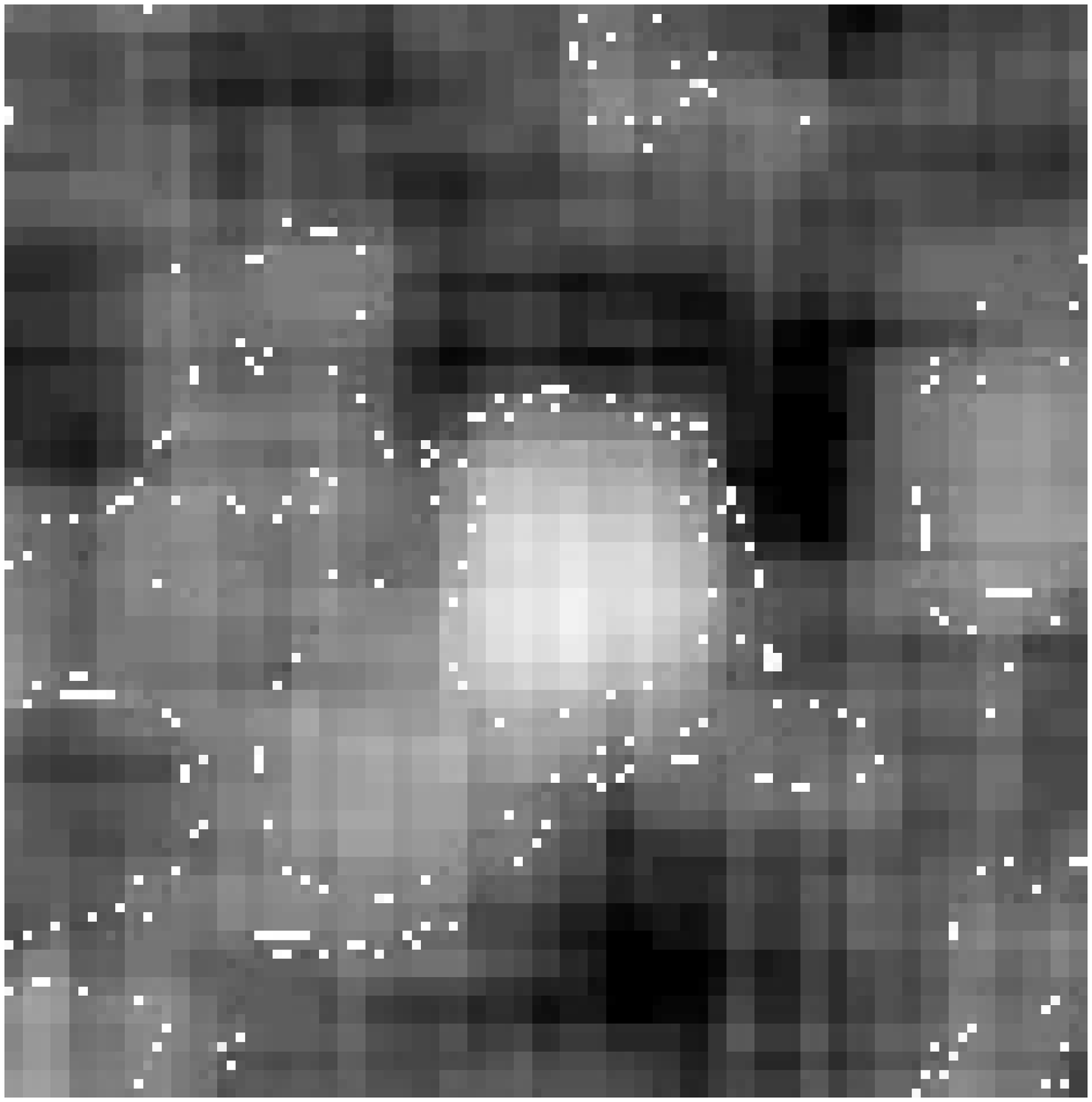}} \\
\fbox{\includegraphics[height=4cm,bb=0  0 610 700]{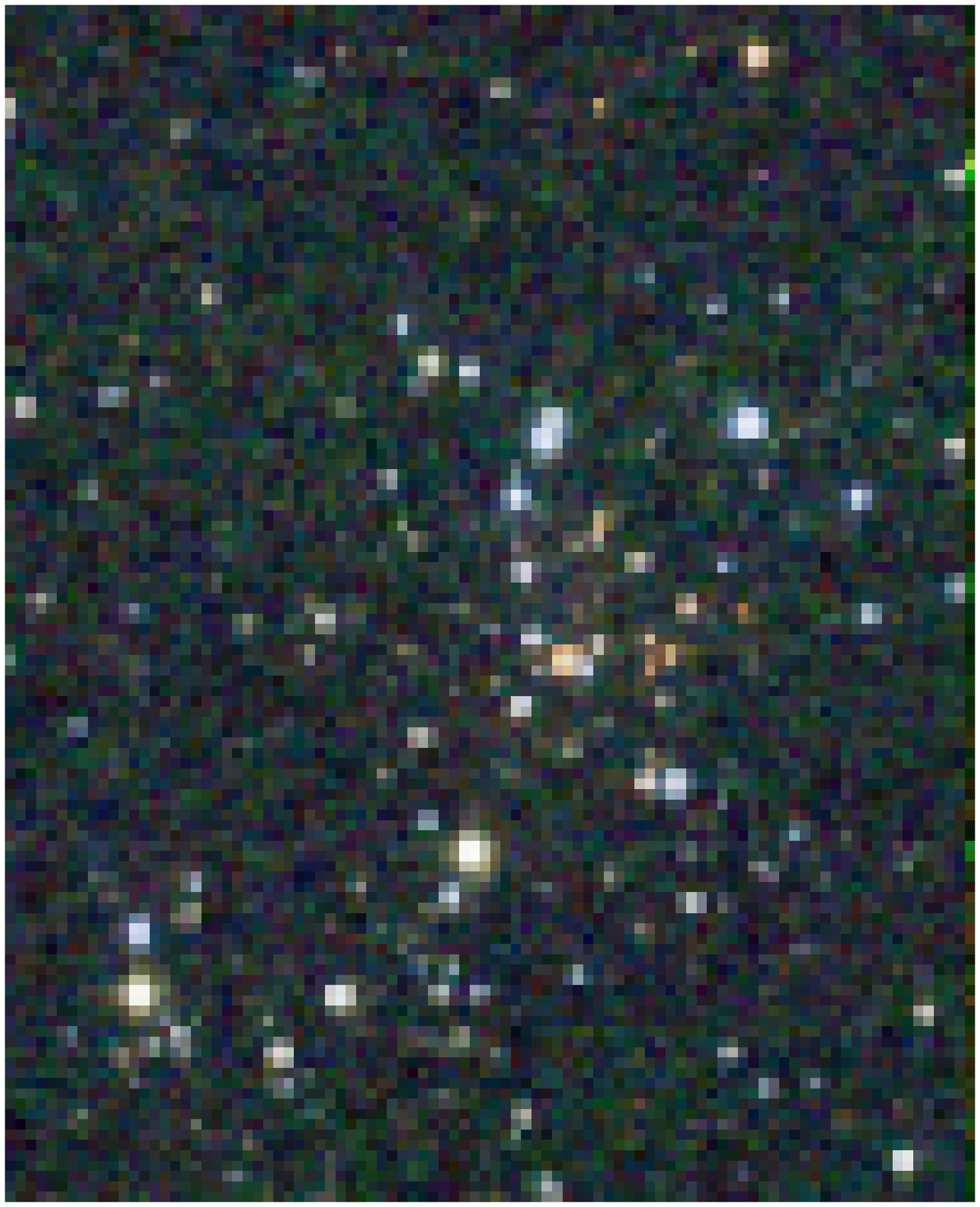}}
\fbox{\includegraphics[height=4cm,bb=0  0 720 720]{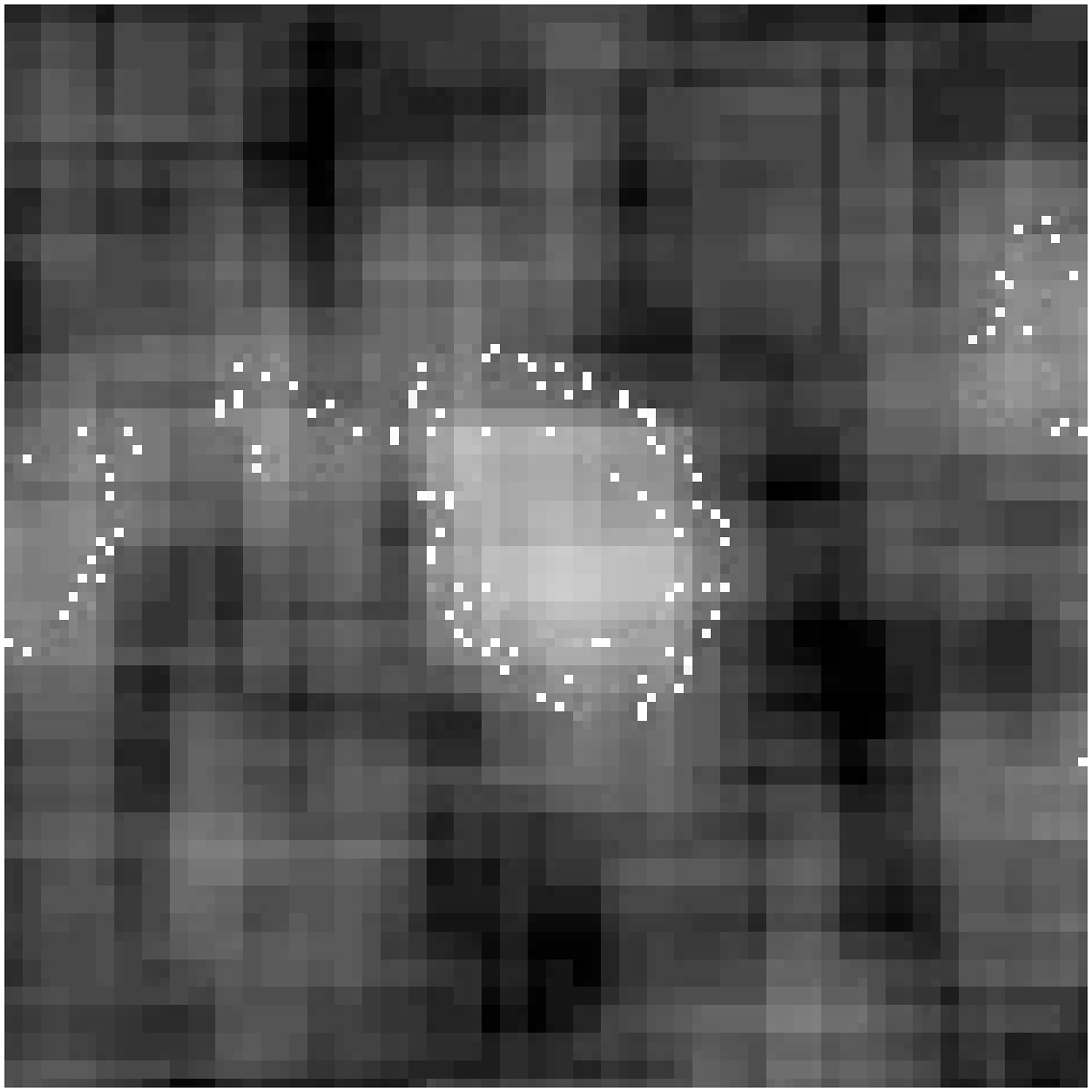}}
\fbox{\includegraphics[height=4cm,bb=0  0 610 700]{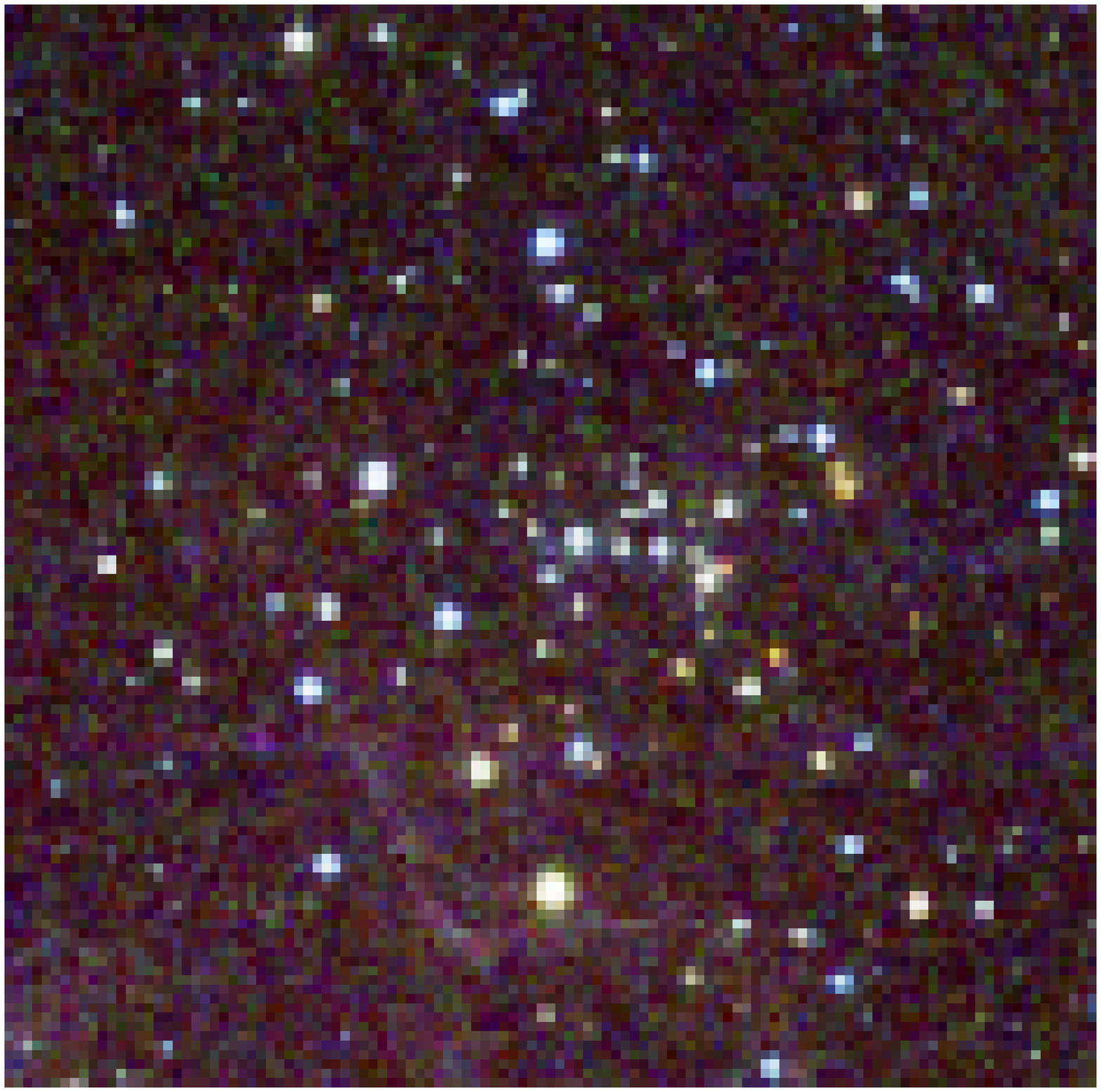}}
\fbox{\includegraphics[height=4cm,bb=0  0 720 720]{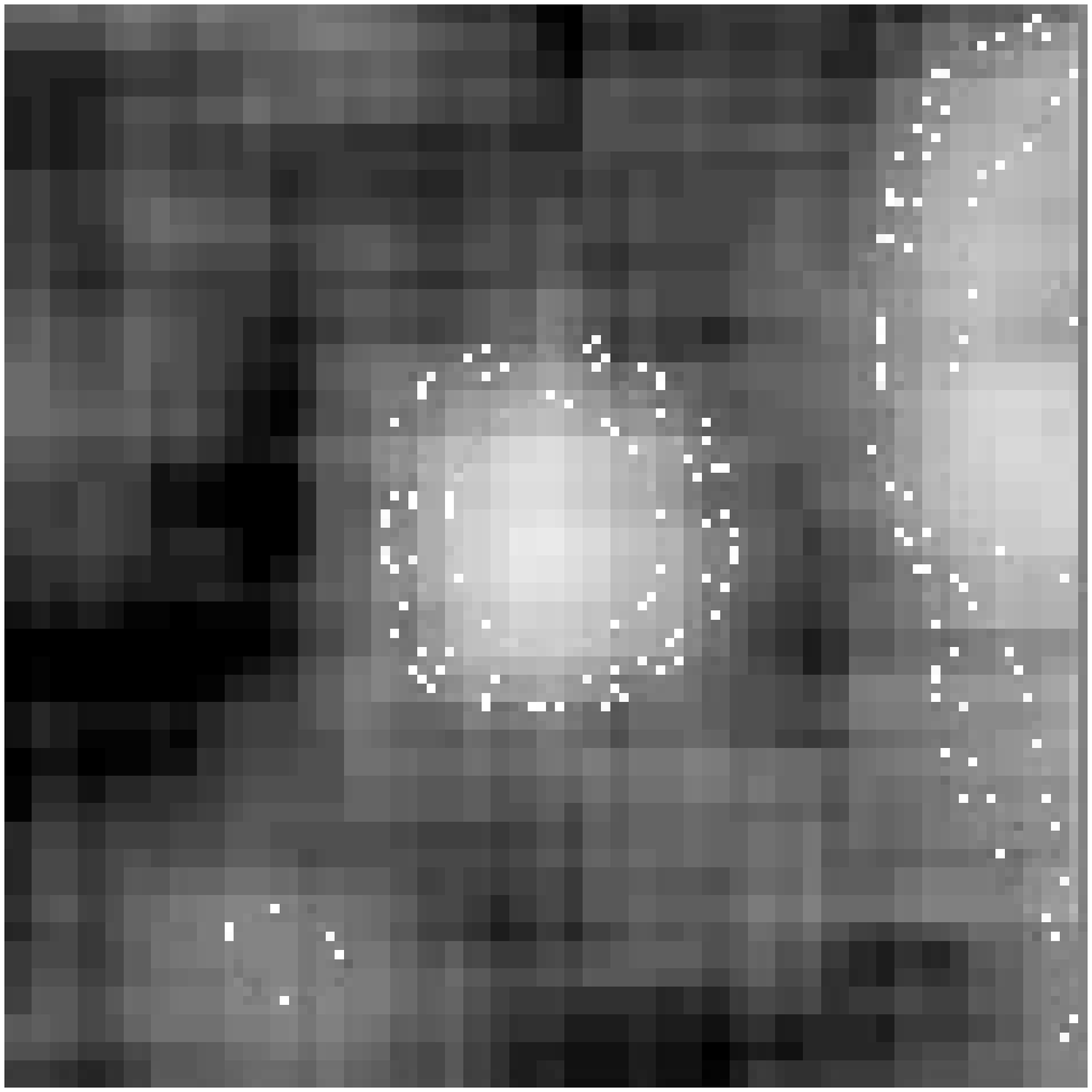}} \\
\fbox{\includegraphics[height=4cm,bb=0  0 610 700]{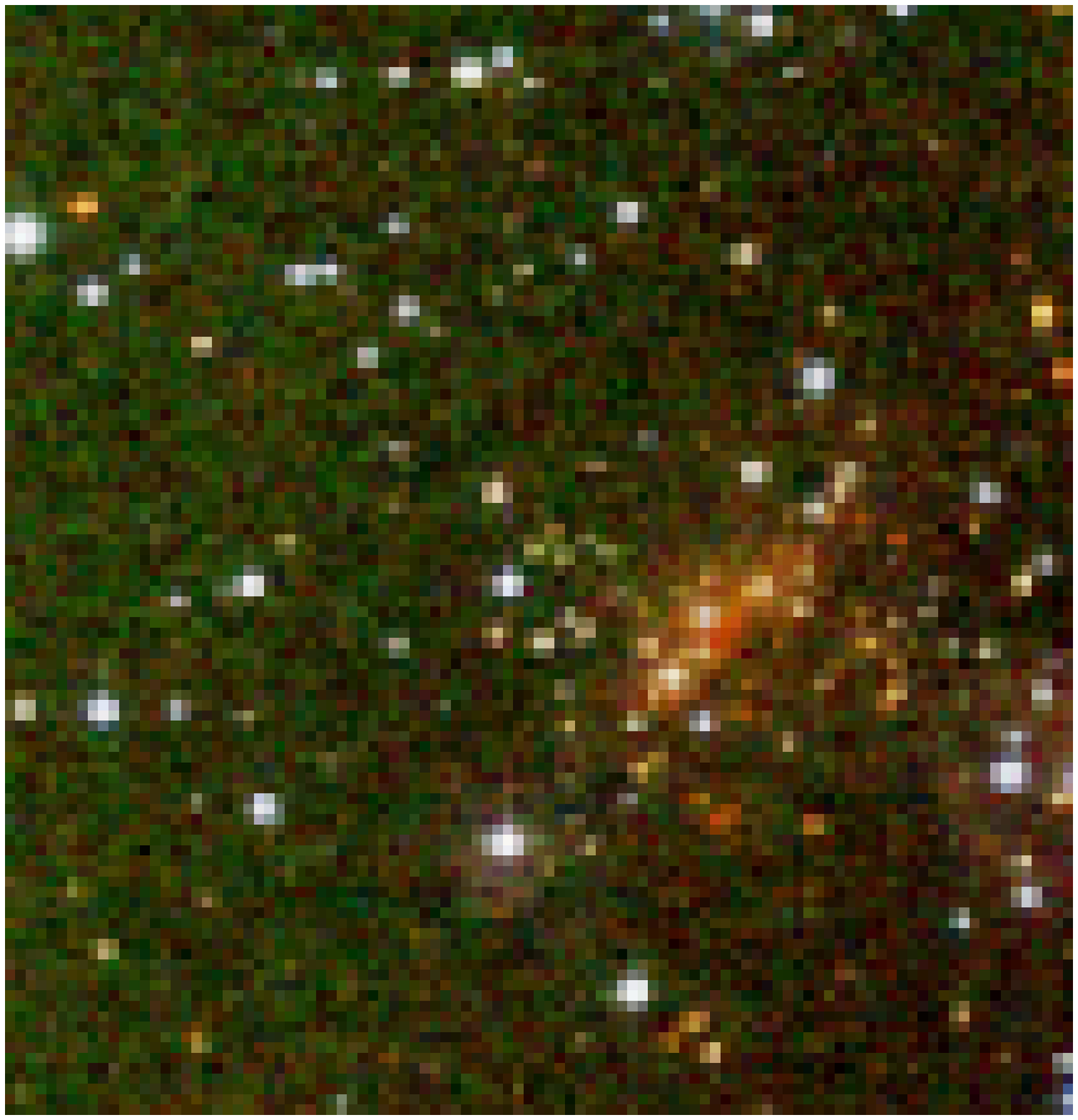}}
\fbox{\includegraphics[height=4cm,bb=0  0 720 720]{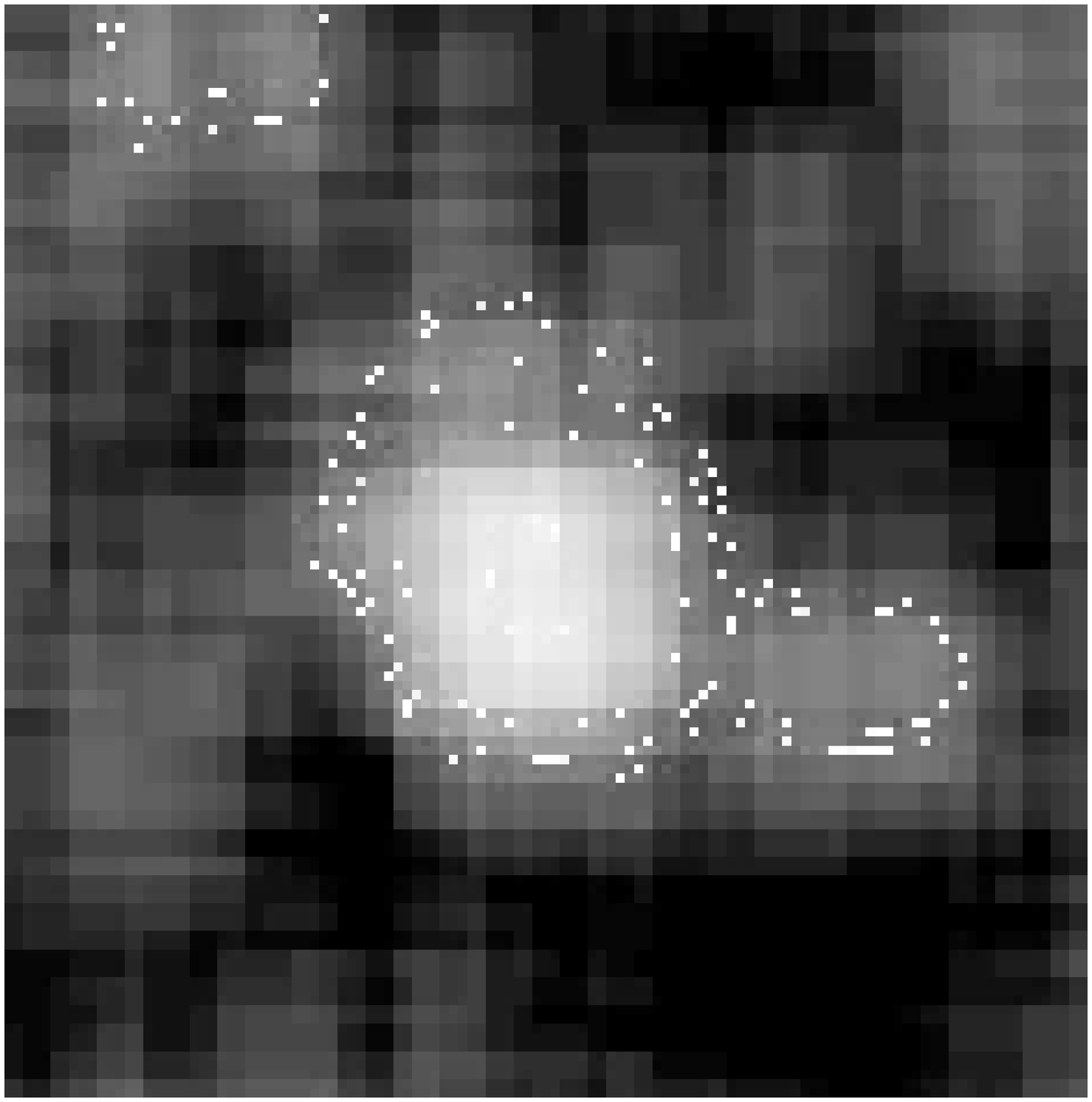}}
\fbox{\includegraphics[height=4cm,bb=0  0 610 700]{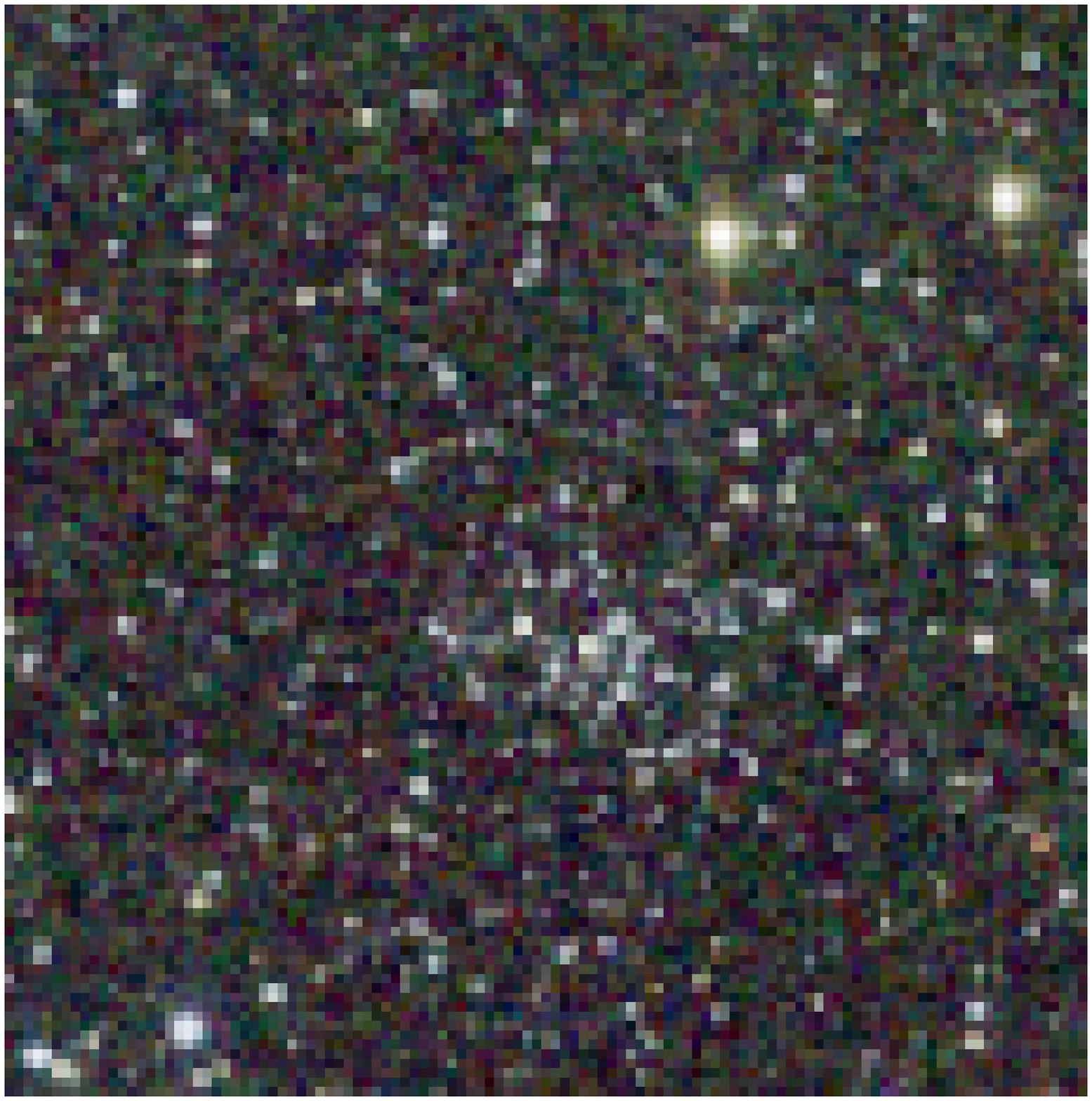}}
\fbox{\includegraphics[height=4cm,bb=0  0 720 720]{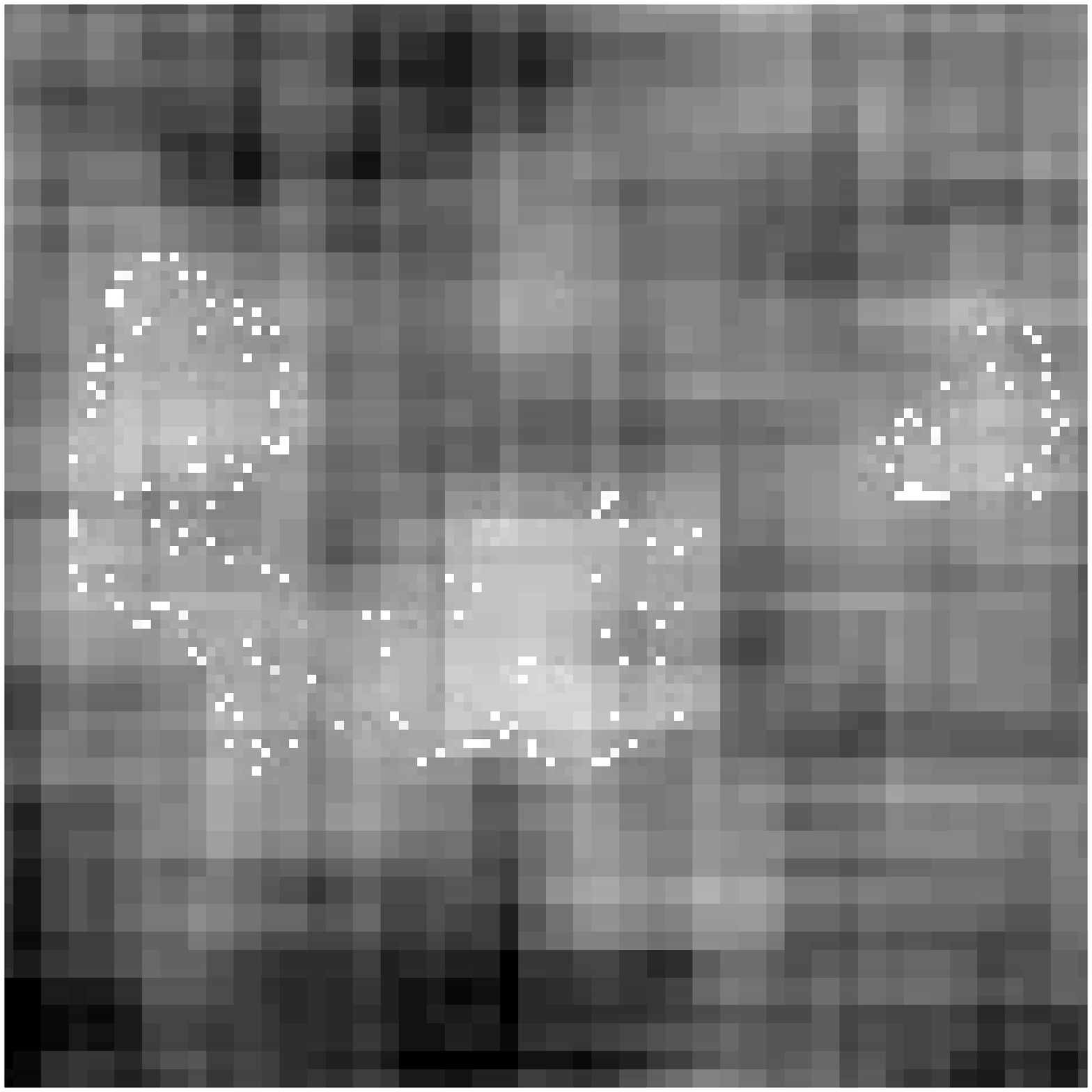}} \\
\caption{\label{twonew} 2MASS JHK colour (left) and K-band star density
(right) maps for the new cluster candidates 0190\,(Qual. Flag\,1), 0191\,(0),
0197\,(1), 0426\,(3), 0488\,(3), 0784\,(4), 0849\,(4), 1476\,(3); from top left
to bottom right. The size of the boxes around the JHK colour images is 5
arcminutes, north is up, east is left. In the star density maps bright regions
indicate enhanced star density. Contrary to the JHK images the K-band star
density maps are in Galactic coordinates, possess a size of 0.25\,degrees and
are centred on the cluster position. Note that in the JHK images the clusters
are not always centred due to the available 2MASS image data.} 
\end{figure*}

Please note that the SIMBAD classification is not 100\,\% reliable for all
individual objects. However, single mis-classifications will not influence the
statistical analysis performed in this paper. One possible reason for such 
mis-classifications are erroneous coordinates of known clusters in the SIMBAD 
database: We found, for example, that the SIMBAD coordinates of the known OpCl 
Berkeley\,51, which is 3.9\,arcmin away from one of our new cluster candidates 
(Cl\,0197, core radius 0.6\,arcmin), are probably inaccurate. An inspection of
the 2MASS images reveals no star cluster like object at the SIMBAD coordinates
of Berkeley\,51, but clearly a star cluster at the coordinates of our
candidate  (see also Fig.\,\ref{twonew}). Some other examples for potentially
erroneous coordinates in the SIMBAD database are the objects Cl\,0191 --
Berkeley\,49; Cl\,0361 -- Berkeley\,98; Cl\,1476 -- VDBH\,63. We have
therefore searched the vicinity (r\,$<$\,10') of all our new cluster candidates
to find such objects. In Tables\,\ref{sourcelist} and \ref{sourcelist_sde} we
provide the names of all known star clusters closer than 10' and outside three
times the candidate core radius. There are 68 (7\,\%) cases of such a close-by
known cluster outside three times the core radius of our candidate. 

In total 1021 of our cluster candidates have no known entry in the SIMBAD
database. In the Appendix in Tables\,\ref{sourcelist} and
\ref{sourcelist_sde} we summarise the properties of our candidates. Note that
these tables will be available online only.

\begin{figure*}
\fbox{\includegraphics[height=4cm,bb=0  0 610 700]{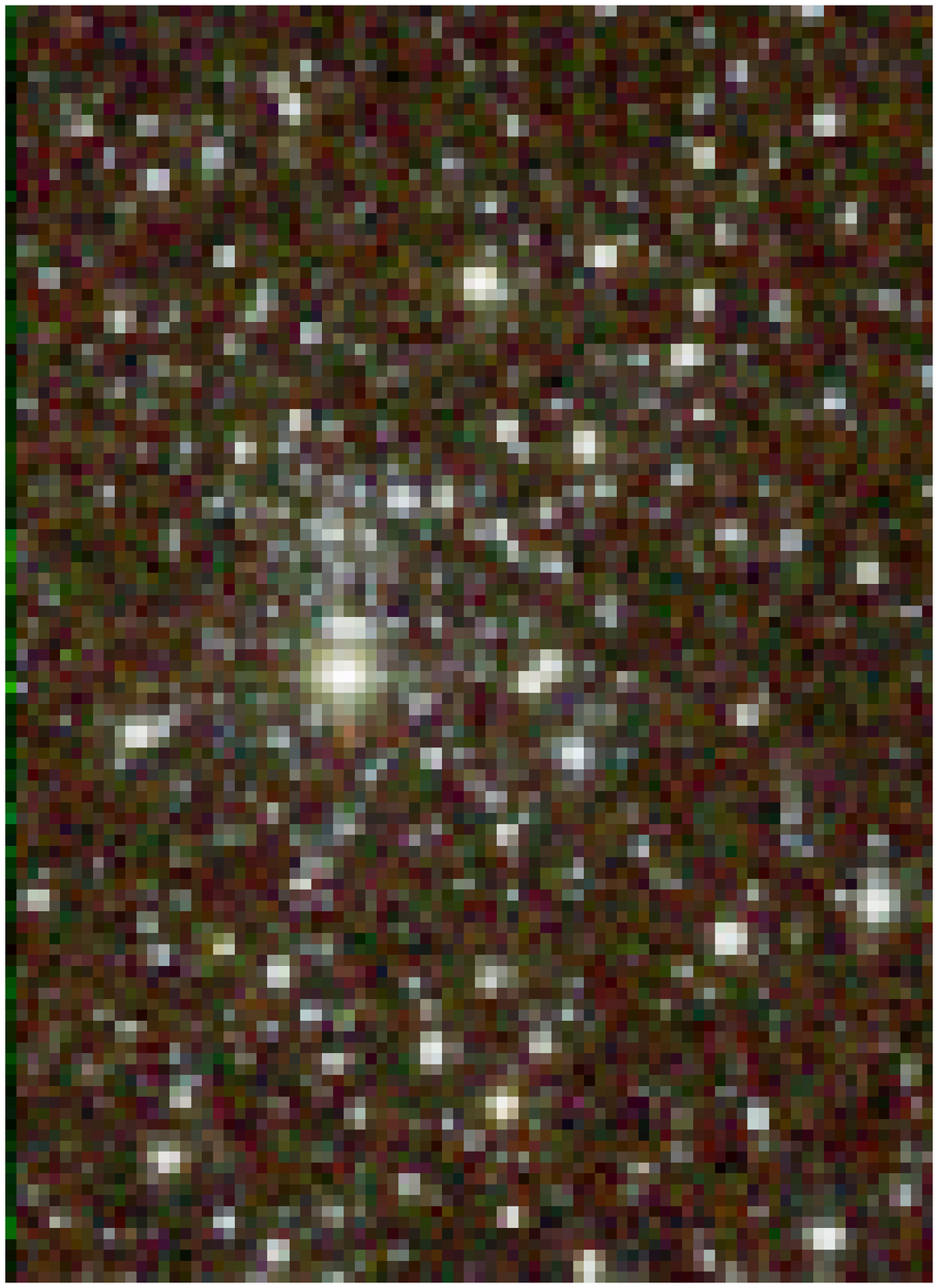}}
\fbox{\includegraphics[height=4cm,bb=0  0 720 720]{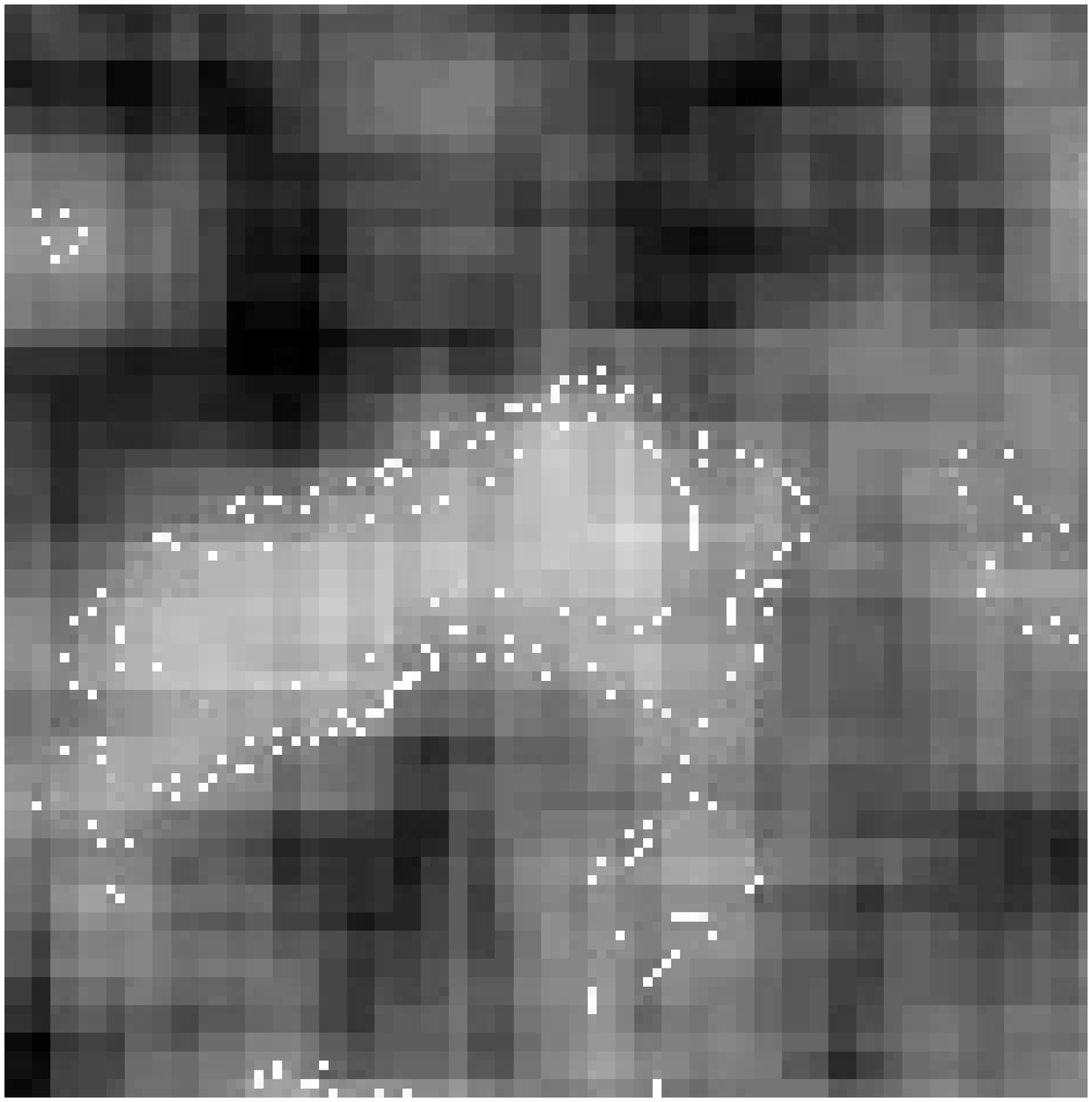}}
\fbox{\includegraphics[height=4cm,bb=0  0 610 700]{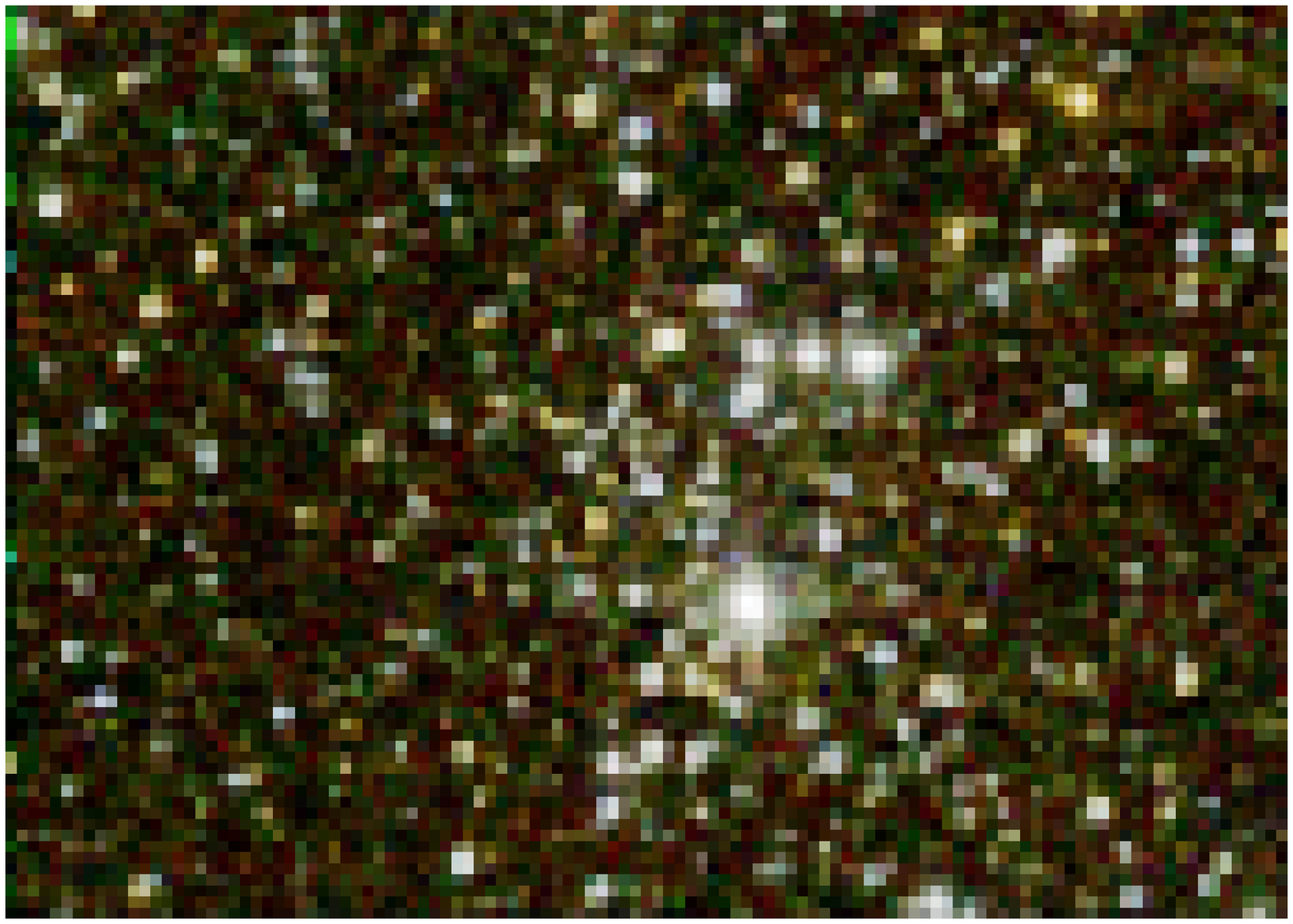}}
\fbox{\includegraphics[height=4cm,bb=0  0 720 720]{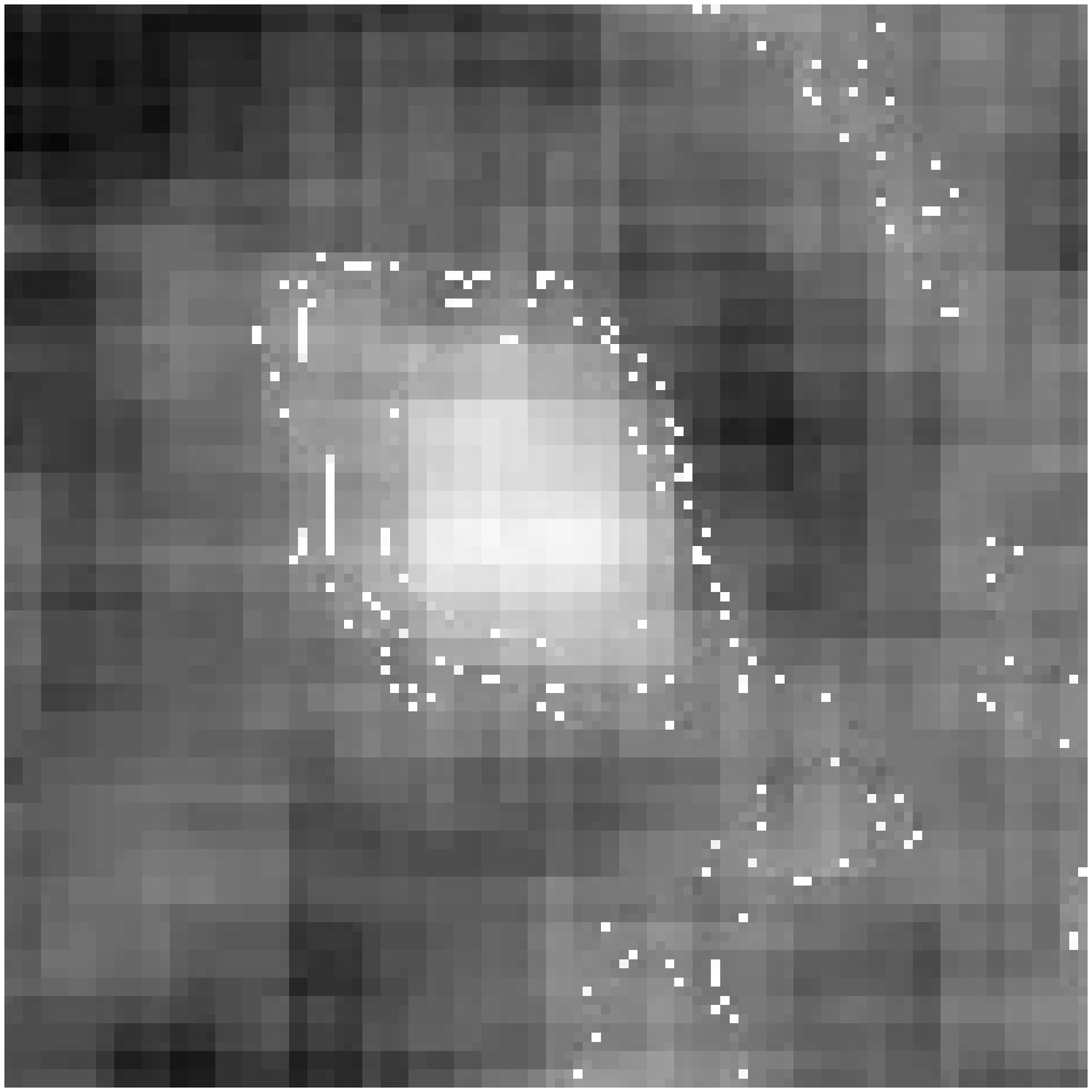}} \\
\fbox{\includegraphics[height=4cm,bb=0  0 610 700]{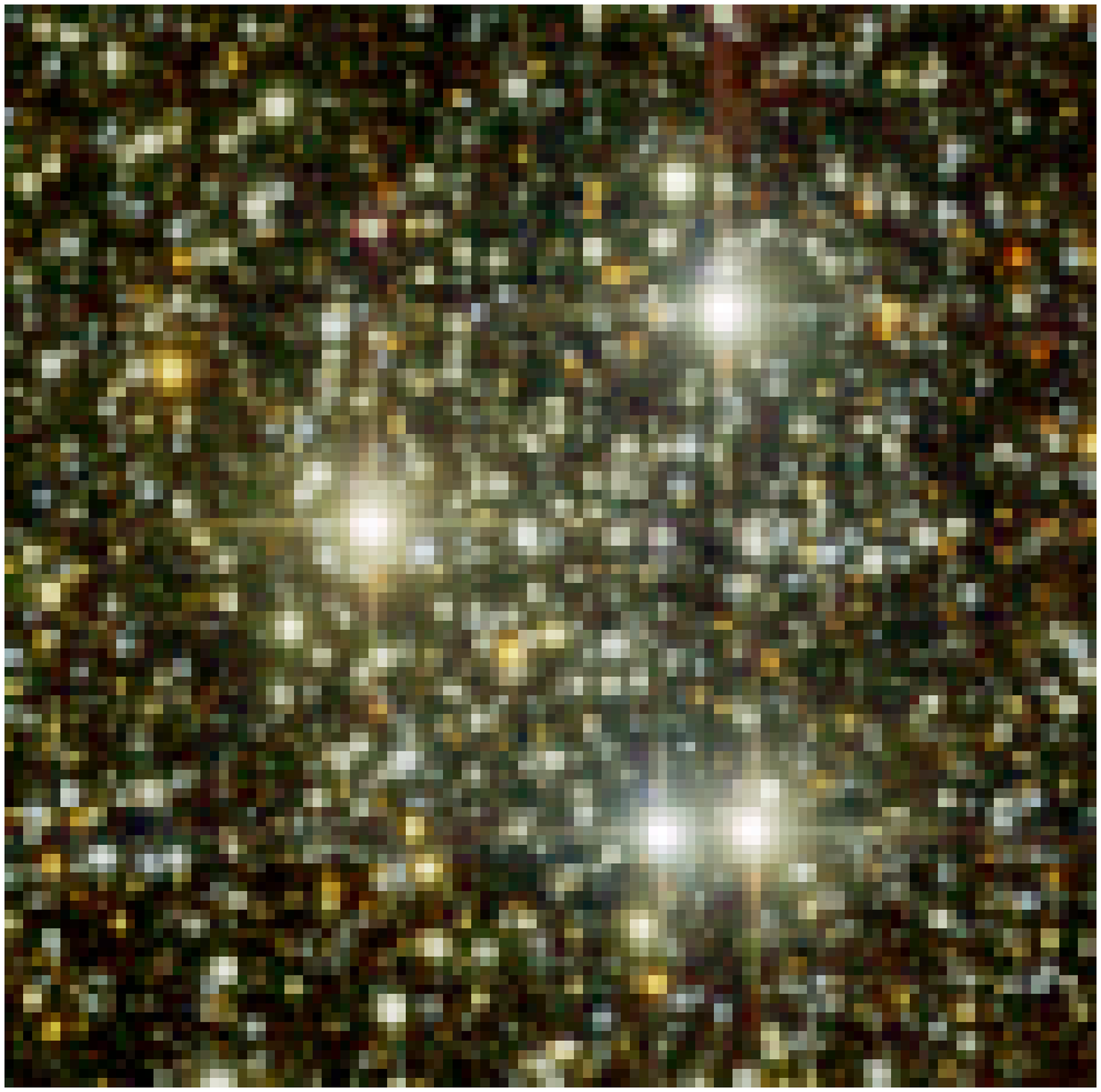}}
\fbox{\includegraphics[height=4cm,bb=0  0 720 720]{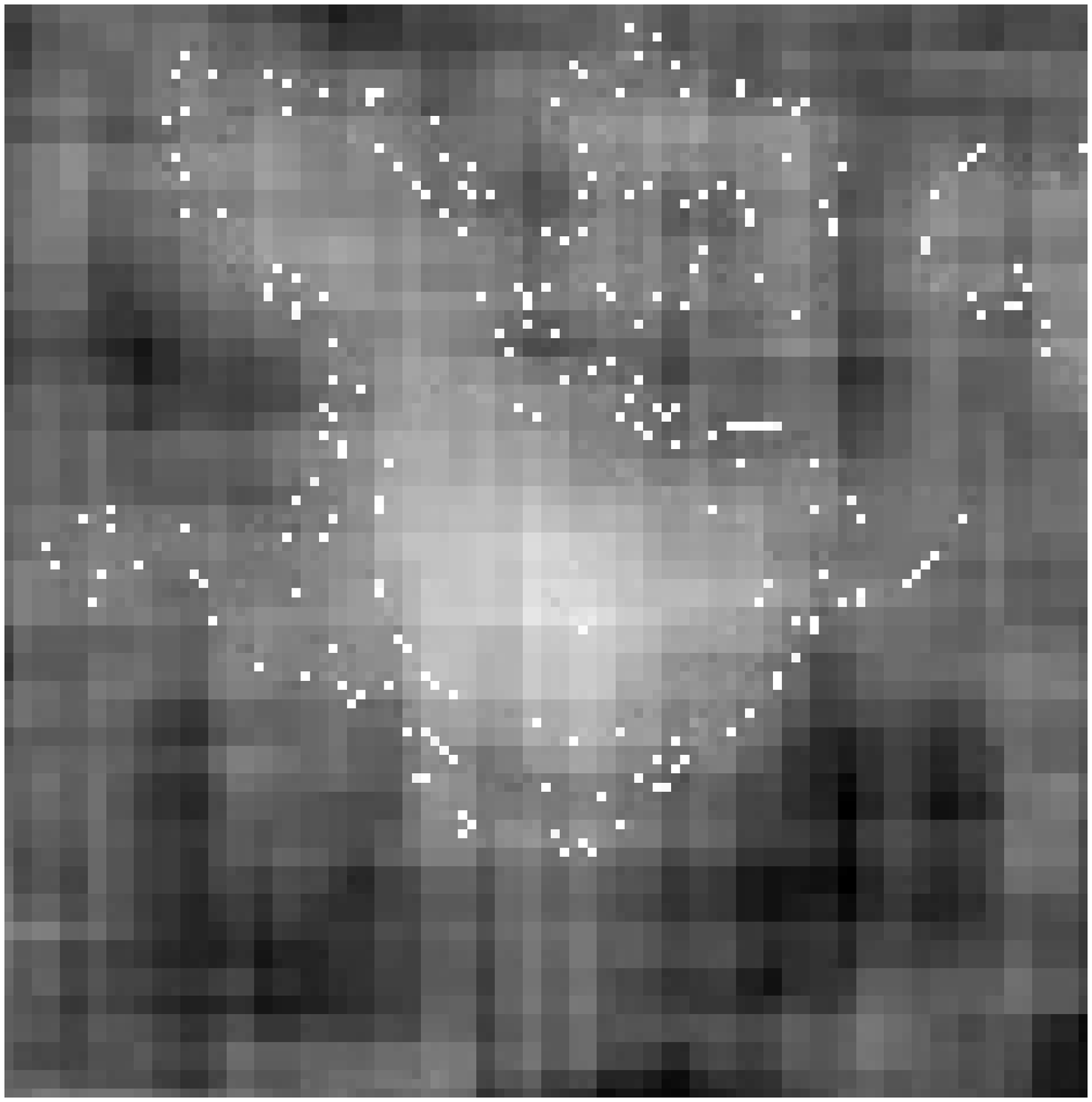}}
\fbox{\includegraphics[height=4cm,bb=0  0 610 700]{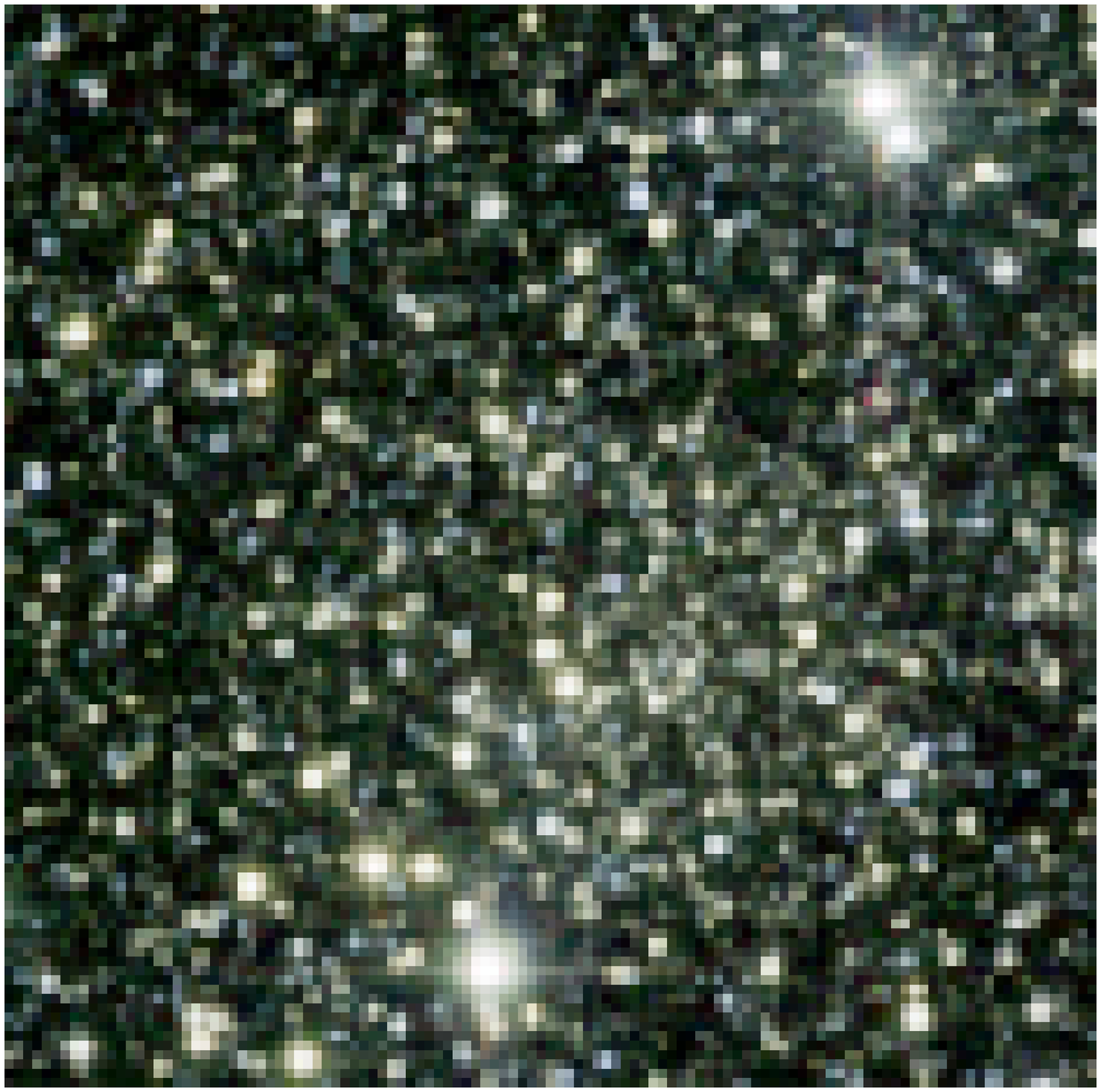}}
\fbox{\includegraphics[height=4cm,bb=0  0 720 720]{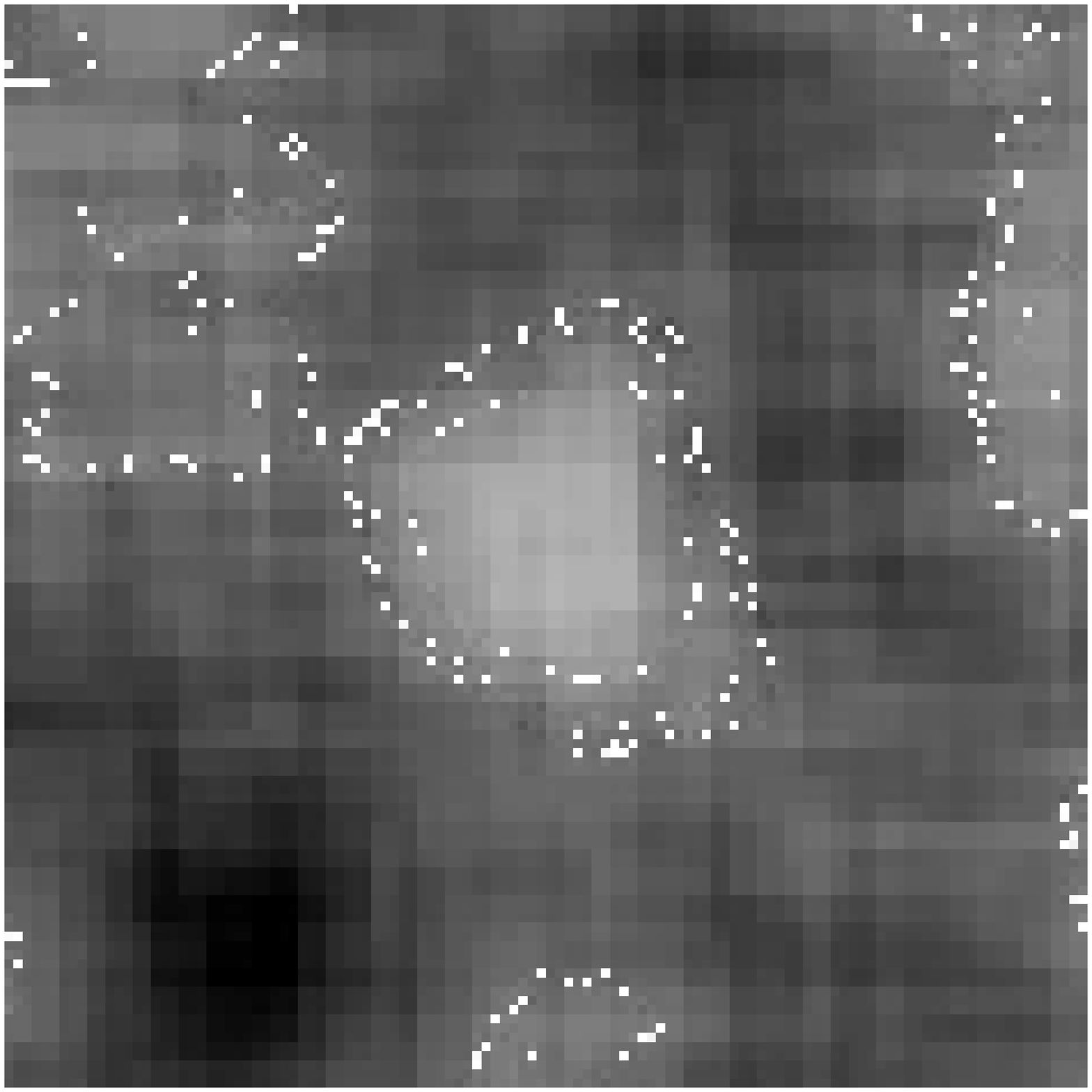}} \\
\fbox{\includegraphics[height=4cm,bb=0  0 610 700]{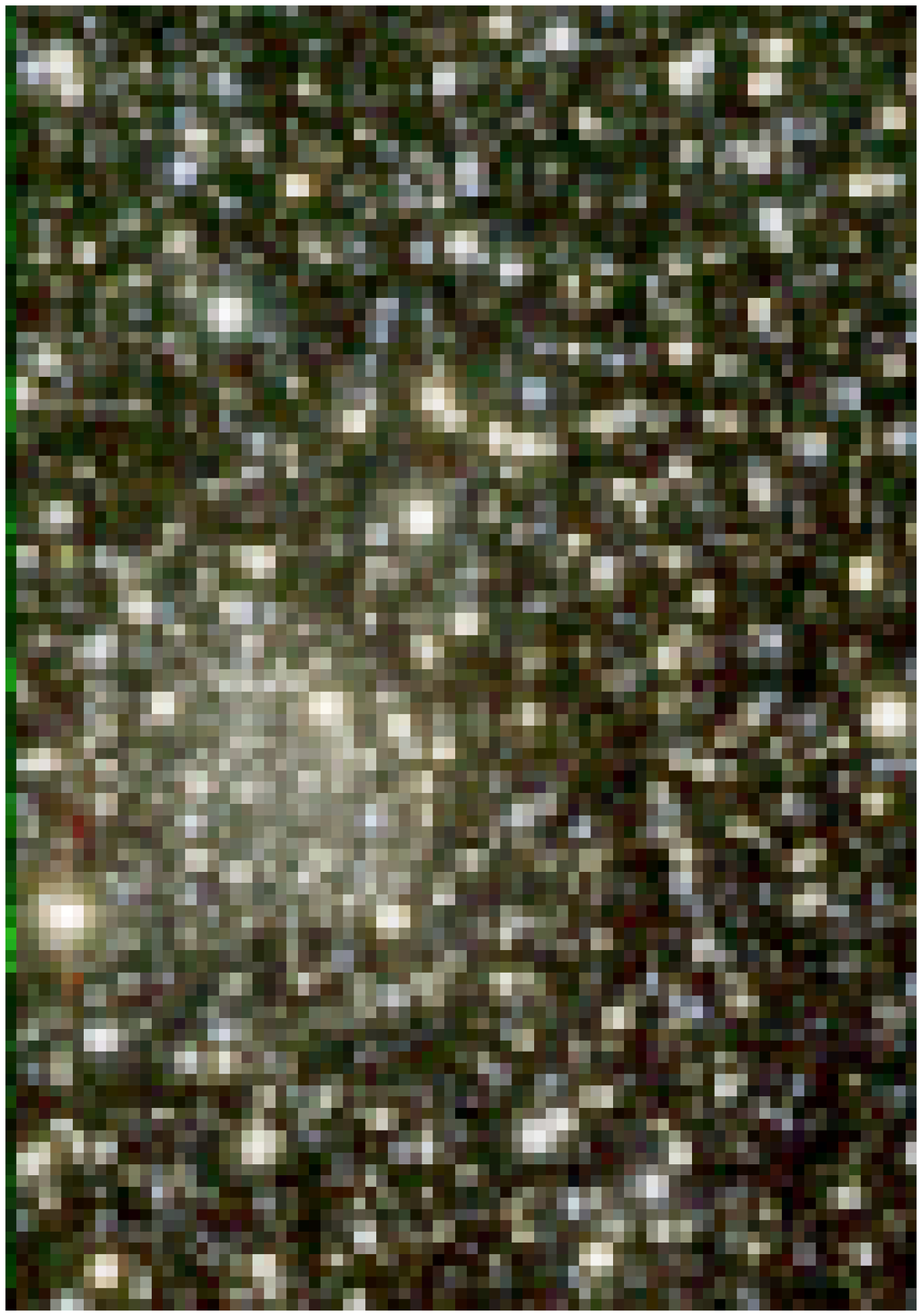}}
\fbox{\includegraphics[height=4cm,bb=0  0 720 720]{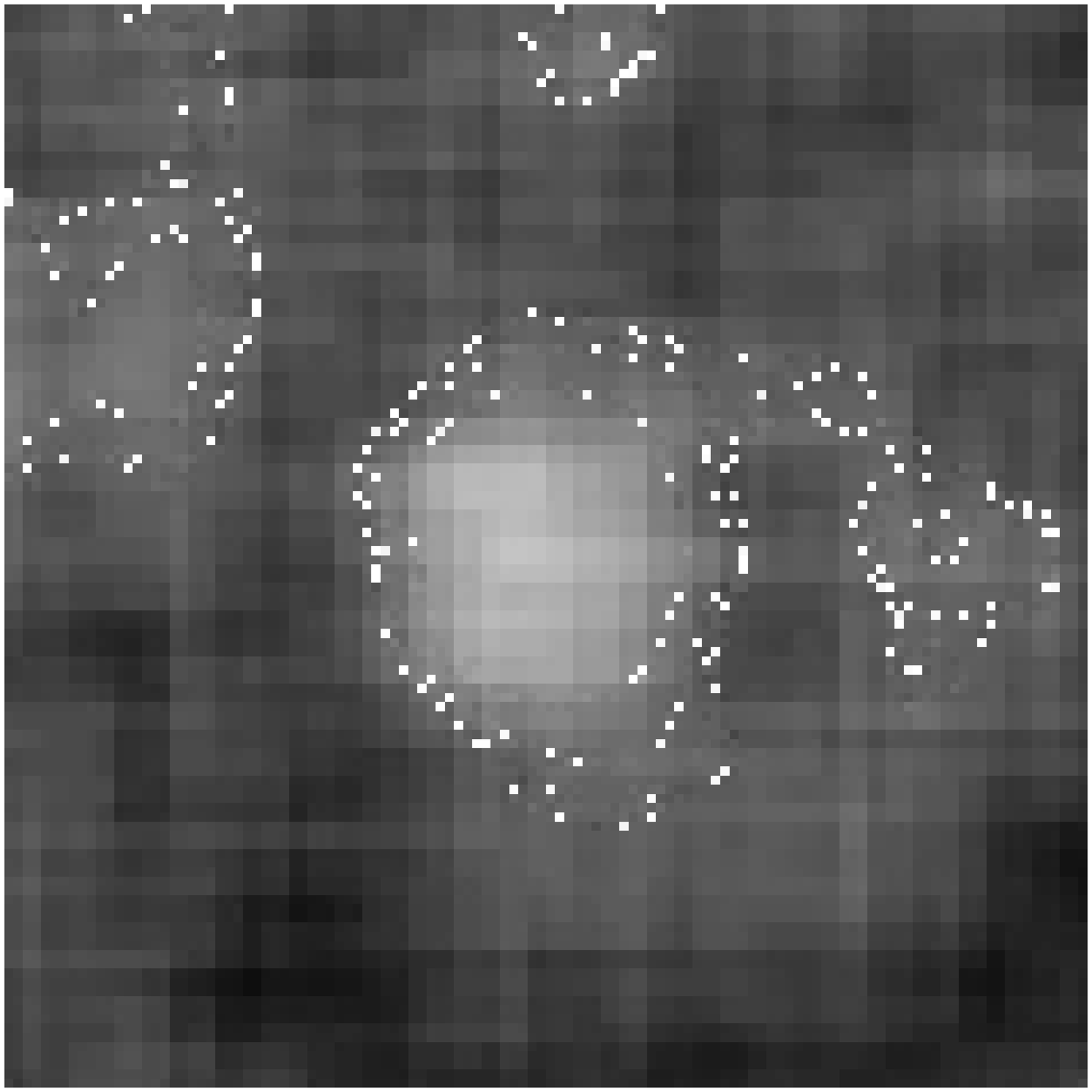}}
\fbox{\includegraphics[height=4cm,bb=0  0 610 700]{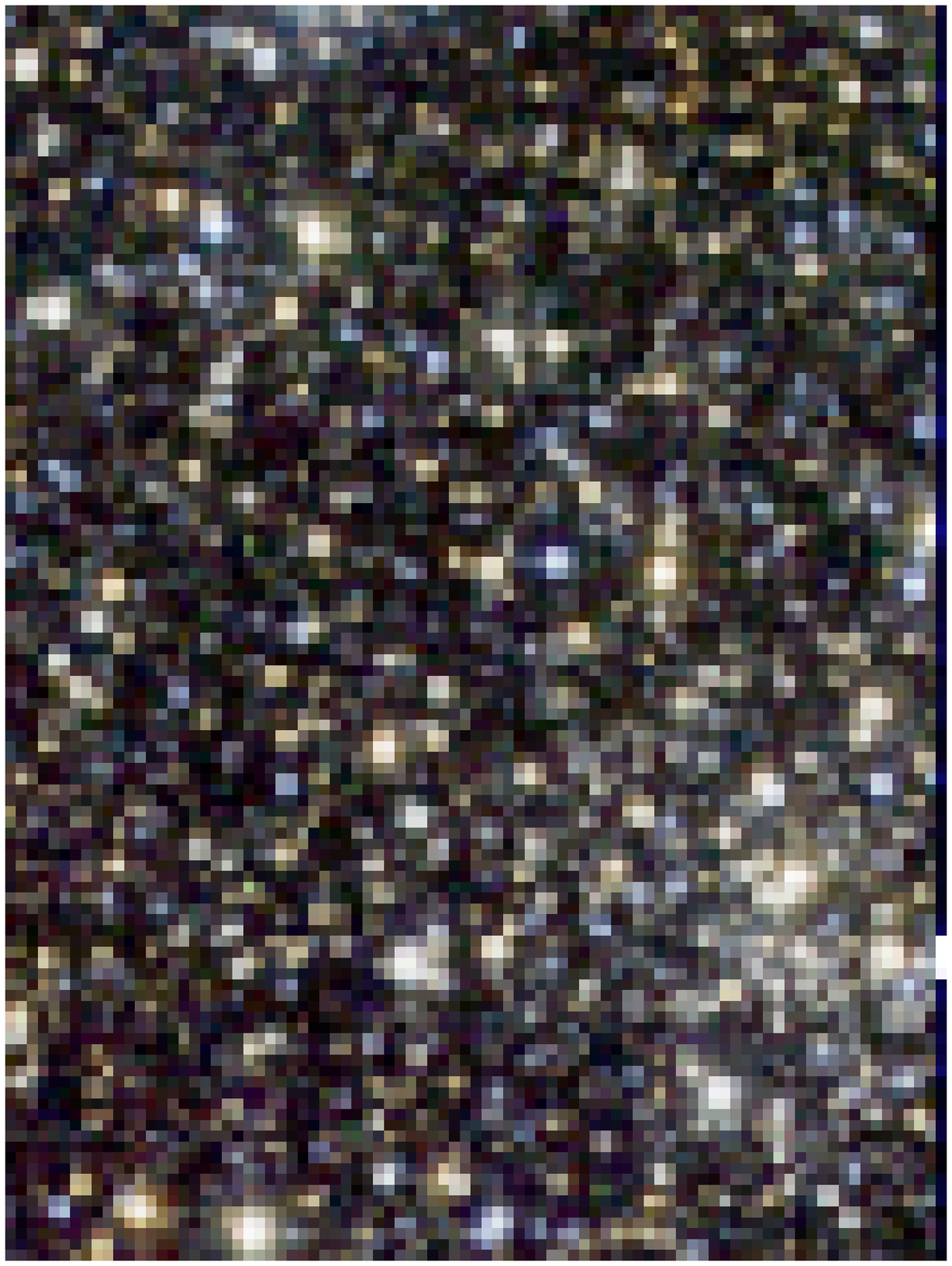}}
\fbox{\includegraphics[height=4cm,bb=0  0 720 720]{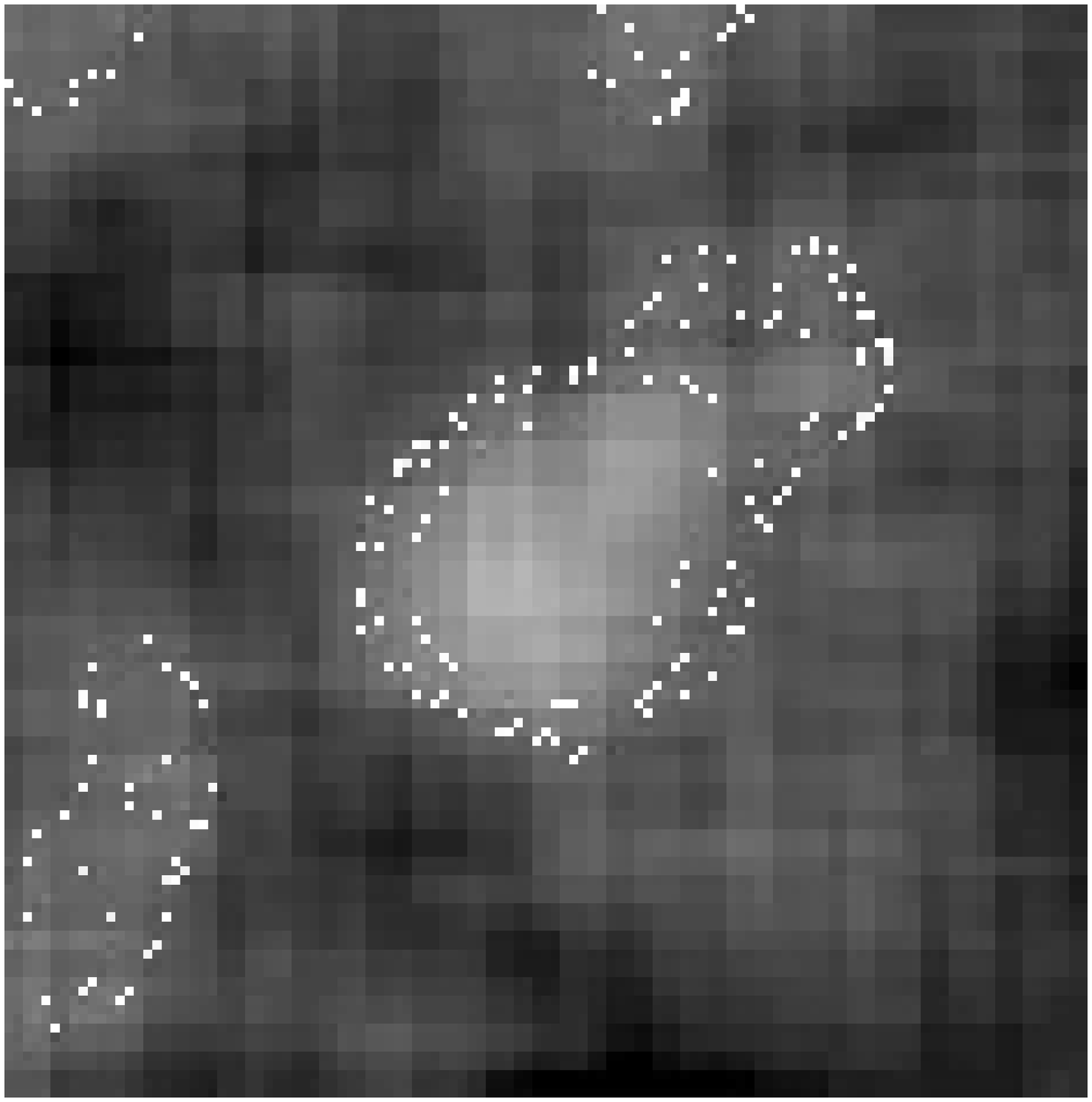}} \\
\caption{\label{twonew2} As Fig.\,\ref{twonew} but for the new cluster
candidates 1483\,(1), 1530\,(2), 1656\,(3), 1716\,(2), 1735\,(4), 1744\,(3);
from top left to bottom right.}
\end{figure*}

\subsection{Reliability of our cluster candidates}
\label{reliab}
  
How reliable are the new cluster candidates, i.e. how significant is the
contamination of our sample with random star density fluctuations of field
stars? Here we will discuss several approaches to quantify the contamination of
our sample statistically. 

\begin{figure*}
\includegraphics[height=17.5cm,angle=-90,bb=50 60 540 780]{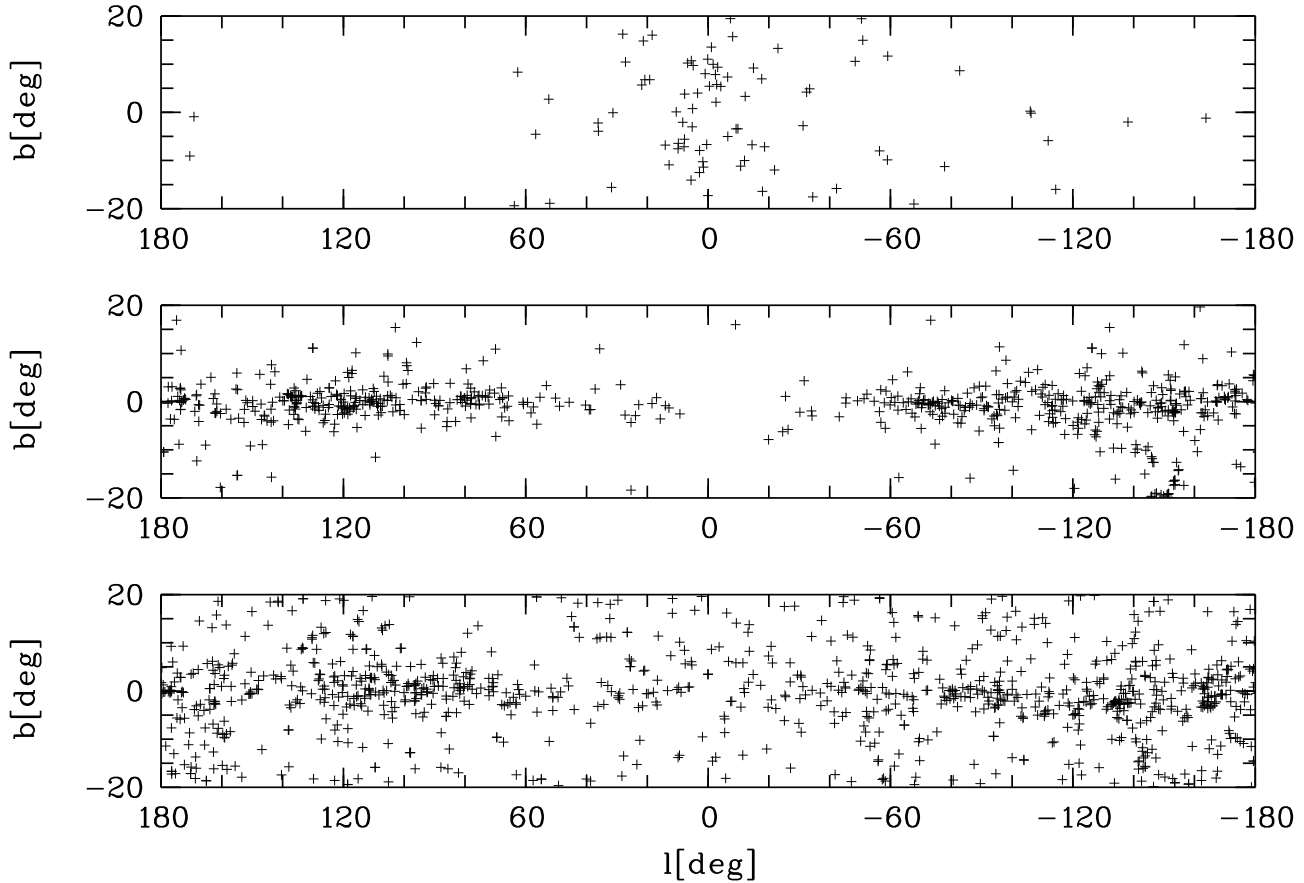}
\caption{\label{spatialdistribution} Distribution of the known clusters and new
cluster candidates in our search area. {\bf Top:} Known GlCls. {\bf Middle:}
Known OpCls. {\bf Bottom:} New star cluster candidates. The different
distributions of the two types of stars clusters can be seen clearly. Also the
lack of known OpCls near the Galactic Centre is evident in the middle panel.}
\end{figure*}

\begin{enumerate}

\item[\bf Spatial Distribution:] We analyse the spatial distribution of our
cluster candidates in Sect.\,\ref{sp_dist}. A comparison of the distributions
of known OpCls and new cluster candidates shows that about 50\,\% of the new
candidates are distributed homogeneously in our map (see left panel of
Fig.\,\ref{spatialdistribution}). Thus they do not follow the distribution of
known OpCl, and are therefore likely contaminating objects (see 
Sect.\,\ref{sp_dist} for details). This provides a first estimate of the 
contamination in our sample of 50\,\%. 

\item[\bf Detection Method:] We detect 73/86 (85\,\%) of the know GlCls and
435/681 (64\,\%) of the known OpCls in our sample automatically. In contrast,
only 274/1021 (27\,\%) of the new candidates are detected automatically. If we
assume that the ratio of automatically/manually detected known OpCls applies
also for the new candidates, we should have found only 428 new clusters. Hence
we can estimate a contamination of (1021-428)/1021\,=\,58\,\%. Since new
clusters are likely to possess few members (because they remained undiscovered
so far) this number can be considered to be an upper limit. Thus, it is in good
agreement  with the above estimated contamination of 50\,\%. It also indicates
that the  automatically detected cluster candidates are less contaminated. We
have marked  all cluster candidates in Tables\,\ref{sourcelist} and
\ref{sourcelist_sde} that are detected automatically.

\item[\bf Cluster Pairs:] In Sect.\,\ref{clustering} we investigate the
probability $P(r)$ to find pairs of clusters with a given separation $r$. If
one subsample of our clusters is dominated by contamination we should clearly
see a difference in the probability distribution for cluster pairs. We
compared $P(r)$ for: a) the known OpCls in our sample; b) all new cluster
candidates; c) the probable cluster candidates; d) the manually detected
clusters; e) the automatically detected clusters. We find that there is no
statistically significant difference between $P(r)$ for all subsamples. This
indicates that the contamination, albeit present, does not influence our
analysis in Sect.\,\ref{clustering}

\item[\bf Cluster Morphology:] A further possibility to determine the
contamination is the morphological analysis of our cluster candidates in
Sect.\,\ref{propertyanalysis}. We fit radial star density (King) profiles to
all identified objects, and analyse the quality of those fits. We find that the
known GlCls give the best fit: 81/86 (94\,\%) have a quality flag better than 3
(see Sect.\,\ref{propertyanalysis} for how this flag is determined). In the
case of the OpCls we have 483/681 (71\,\%) with a good quality flag. For our
new cluster candidates we obtain for 455/1021 (45\,\%) a good fit. Assuming
again the percentage of the OpCls for the new candidates we should have had a
good fit for 726 objects. This indicates a contamination of
(1021-726)/1021\,=\,29\,\%, somewhat less than the estimates above. In
Tables\,\ref{sourcelist} and \ref{sourcelist_sde} we list for each of the new
cluster candidates the quality flag.

\item[\bf Visual Inspection:] We performed a visual inspection of 2MASS images
for a number of randomly selected new cluster candidates. We inspected JHK
colour images of 60 cluster candidates and selected all objects where an
obvious cluster of stars can be identified visually. We find that for objects
with a quality flag better than 4, about 30\,\% of the cluster candidates show
a clear cluster like appearance in the 2MASS images. To exemplarily demonstrate
the appearance of our candidates in 2MASS images, we present in
Figs.\,\ref{twonew} and \ref{twonew2} JHK colour images and for comparison
K-band star density maps for a  number of candidates, where a clear clustering
of stars is apparent in the JHK image. The 30\,\% of clear cluster candidates
would indicate a contamination of 70\,\%. We note, however, that the visual
impression in images can be misleading if the cluster consists of many faint
stars. Hence the star density maps are a better guide for cluster
identification. Similarly, an increased number of bright stars in an image
might lead to the impression of a cluster, even if the star density map shows
no peak.

\end{enumerate}

In summary with three independent estimates, we find contamination rates
between 30 and 60\,\%, and thus probably around half of our newly identified
candidates are no real clusters. Since we account for this in our statistical
analysis in Sect.\,\ref{clustering}, it does not influence our results. For
individual clusters the determined quality flag, based on statistical estimates
(see Sect.\,\ref{propertyanalysis}), can be used to judge the likelihood that
the  candidate is a real cluster. To verify for the nature for each individual
clusters requires the analysis of colour-magnitude diagrams and thus the
identification of cluster members. Since we concentrate here on the statistical
analyses of our cluster  sample, this is beyond the scope of this work and will
be addressed in a future paper. 

However, using our quality flag and the position of the cluster candidates, we
have identified the most probable new star clusters in our sample. In
particular we selected all cluster candidates that have $|b|<4^\circ$ and
possess a quality flag of less than five (73\,\% of the known OpCls fulfill
this condition). These are 473 objects, their properties are listed in
Table\,\ref{sourcelist}, and we will refere to them as probable clusters. The
majority of the remaining objects are probably only local star density
enhancements, their properties are listed in Table\,\ref{sourcelist_sde}, and
they are referred to as possible clusters.

\section{Overall distribution of star clusters}
\label{overall}

\subsection{Distribution, contamination, completeness}
\label{sp_dist}

\begin{figure*}
\includegraphics[height=8.5cm,angle=-90,bb=30 35 550 800]{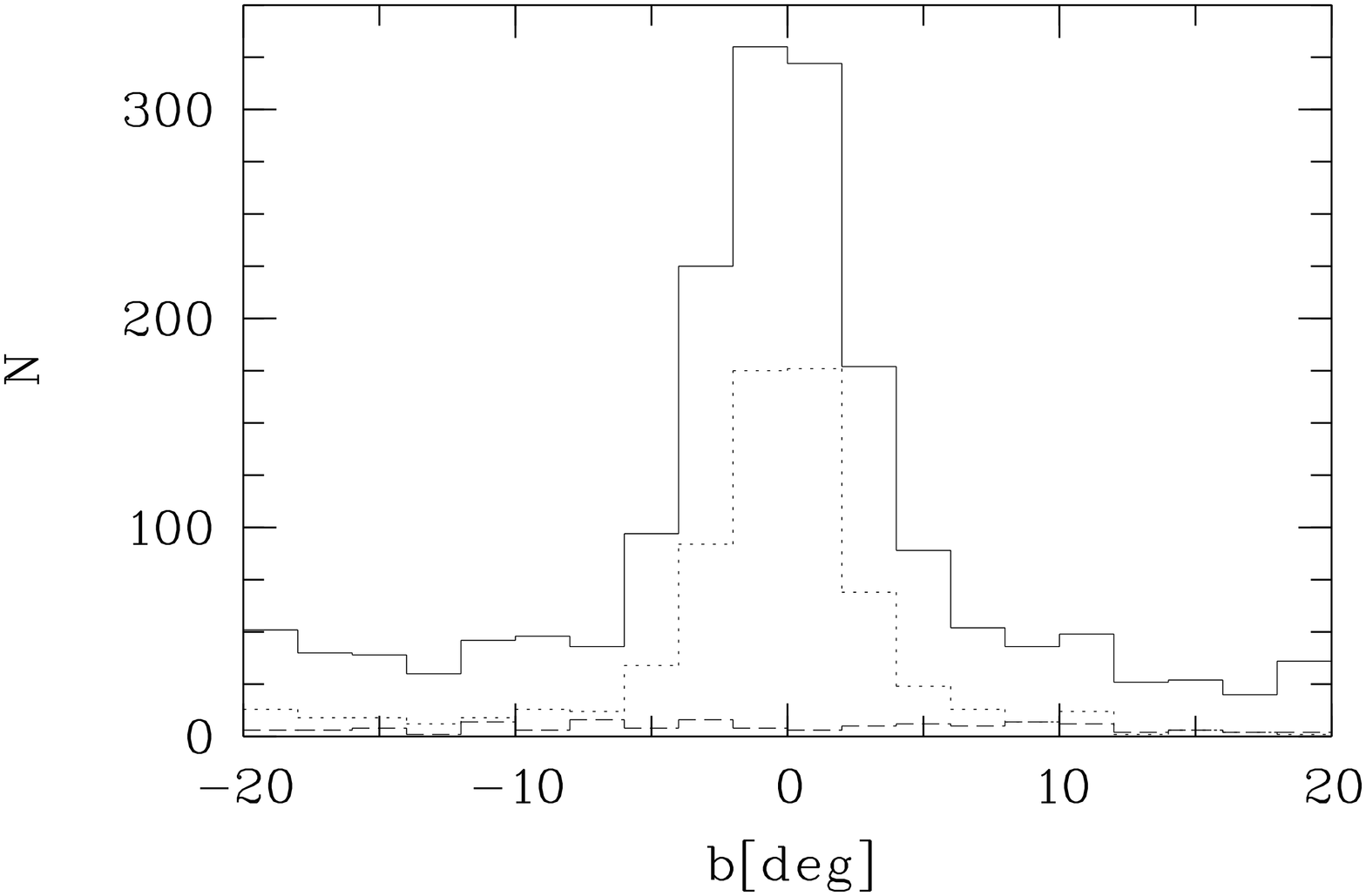}
\hfill
\includegraphics[height=8.5cm,angle=-90,bb=30 35 550 800]{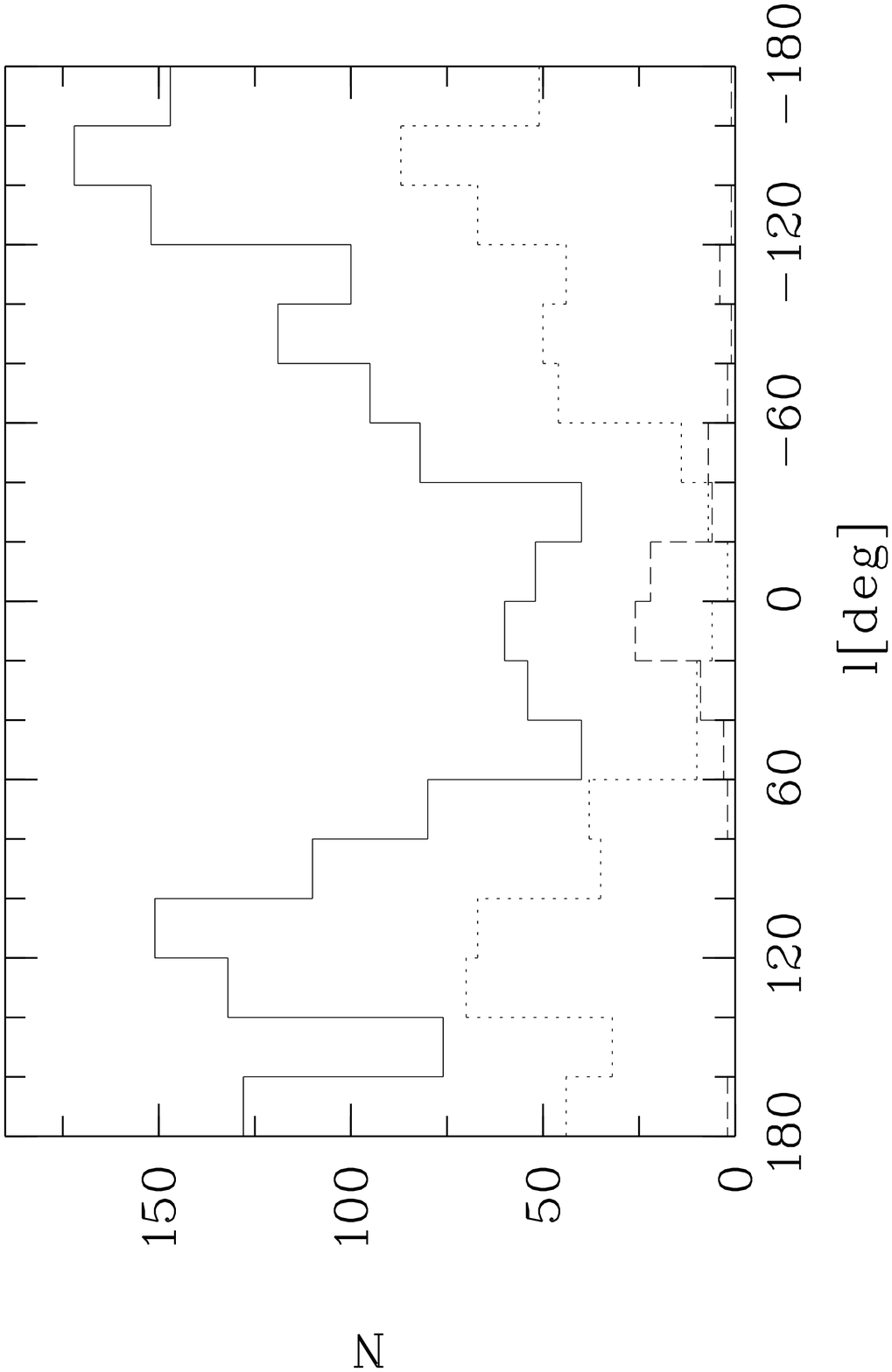}
\caption{\label{histolb} Distribution of our star cluster sample perpendicular
(left panel) and along (right panel) the Galactic Plane. The solid line shows
the histograms for all our objects. The dotted line represents the known OpCls
and the dashed line the known GlCls. Again the apparent lack of clusters toward
the Galactic Centre is evident.}
\end{figure*}

In Fig.\,\ref{spatialdistribution} we plot the distribution of our clusters in
the search area. The upper panel shows the known GlCls contained in our sample.
The concentration towards the Galactic Centre is seen clearly. The middle panel
shows the known OpCls identified in our sample. As expected, this type of
cluster is concentrated towards the Galactic Plane and the main star forming
regions. However, there is a clear lack of objects towards the Galactic Centre
indicative of the fewer known OpCls in this area, as well as a selection effect
of our cluster detection method (evident also in the right panel of
Fig.\,\ref{histolb}). In the lower panel of Fig.\,\ref{spatialdistribution} the
new cluster candidates of our sample are shown. They follow the principle
distribution of the OpCls, reflecting the fact that most of the new candidates
are expected to be OpCls. This panel also shows that there is a component of
about 500 objects which are homogeneously distributed. This population is also
evident in the histogram of the cluster distribution across the Galactic Plane
(left panel of Fig.\,\ref{histolb}). Assuming that the distribution of the
known OpCls is representative, we can conclude that those 500 cluster
candidates are most likely not star clusters, but only local star density
enhancements.  Thus, the contamination of our list of new clusters is likely to
be about 50\,\%, implying that $\sim 500$ of our new cluster candidates are
real, as  already stated in Sect. \ref{reliab}. We note the contaminating
objects, which are homogeneously distributed, do not influence the statistical
analysis of the clustering properties in Sect.\,\ref{clustering}.

By counting the cluster candidates in bins of 60$^\circ$ length along the
Galactic Plane we can estimate the number of potentially missing objects. In
the bins beside the Galactic Centre we find 150 and 180 clusters, respectively.
The region $180^\circ < l < 240^\circ$ possesses the highest number of clusters
(470), while the remaining three bins contain about 320 objects, each. The bin
with 470 clusters is close to the star forming regions of Orion, Taurus, and
Perseus and might represent a local exception from the overall cluster density.
Nevertheless, there are about 300 clusters missing in the area $\pm 60^\circ$
around the Galactic Centre. These clusters are not detected due to the high
background star density (low density contrast between star cluster and
background) and hence provide a promising target area for future projects like
the UKIDDS survey.

This shows that our detection method is subject to selection effects. Only star
clusters that possess a significant density enhancement towards the centre are
detected. As a result, the high star densities towards the Galactic Plane and 
especially towards the Galactic Centre will hamper our detection rate. Hence
any statistical analysis of OpCl properties will have to be constrained to
regions outside $\pm 60^\circ$ from the Galactic Centre. Clusters with few
stars and clusters that are spread out over a large area on the sky are not
detected either. Those types of clusters, however, are not expected to be
distributed in a way different from the detected objects, and hence will not
influence our  analysis in Sect.\,\ref{clustering}.

\subsection{Clustering of clusters}
\label{clustering}

Here we investigate the projected distribution of the star clusters in our
sample. In particular, we are interested in modelling this two dimensional
distribution using simple assumptions. To quantify the distribution of clusters
we determined the probability $P(r)$ to find pairs of clusters with a given
separation $r$. This is done by counting the number of clusters $\mathcal
N_i(r)$ in a ring with a radius $r$ and a thickness $\delta r$ for each cluster
$i$. This has to be normalised by the total number of clusters ($\mathcal N -
1$) and the area of the ring $2 \pi r \delta r$. These individual probabilities
$P_i(r)$ are then averaged to obtain the probability $P(r)$ as follows:
\begin{equation}
P(r) = \frac{1}{\mathcal N} \cdot \sum\limits_{i=1}^{\mathcal N} P_i(r) =
\frac{1}{{\mathcal N} \cdot ({\mathcal N - 1})} \cdot
\sum\limits_{i=1}^{\mathcal N} \frac{\mathcal N_i(r)}{2 \cdot \pi \cdot r \cdot
\delta r}  
\label{prob_eq}
\end{equation}

Note that we used $\delta r = 3'$ as thickness of the ring. Since we
encountered a significant drop in the number of detected clusters towards the
Galactic Centre ($\pm 60^\circ$), we exclude this area from the analysis. This
leaves $\mathcal N = 1461$ cluster candidates. There are only 13 known GlCls in
this subsample, as well as 632 known OpCls and 816 new candidates.

\begin{figure*}
\includegraphics[height=8.5cm,angle=-90,bb=20 30 550 770]{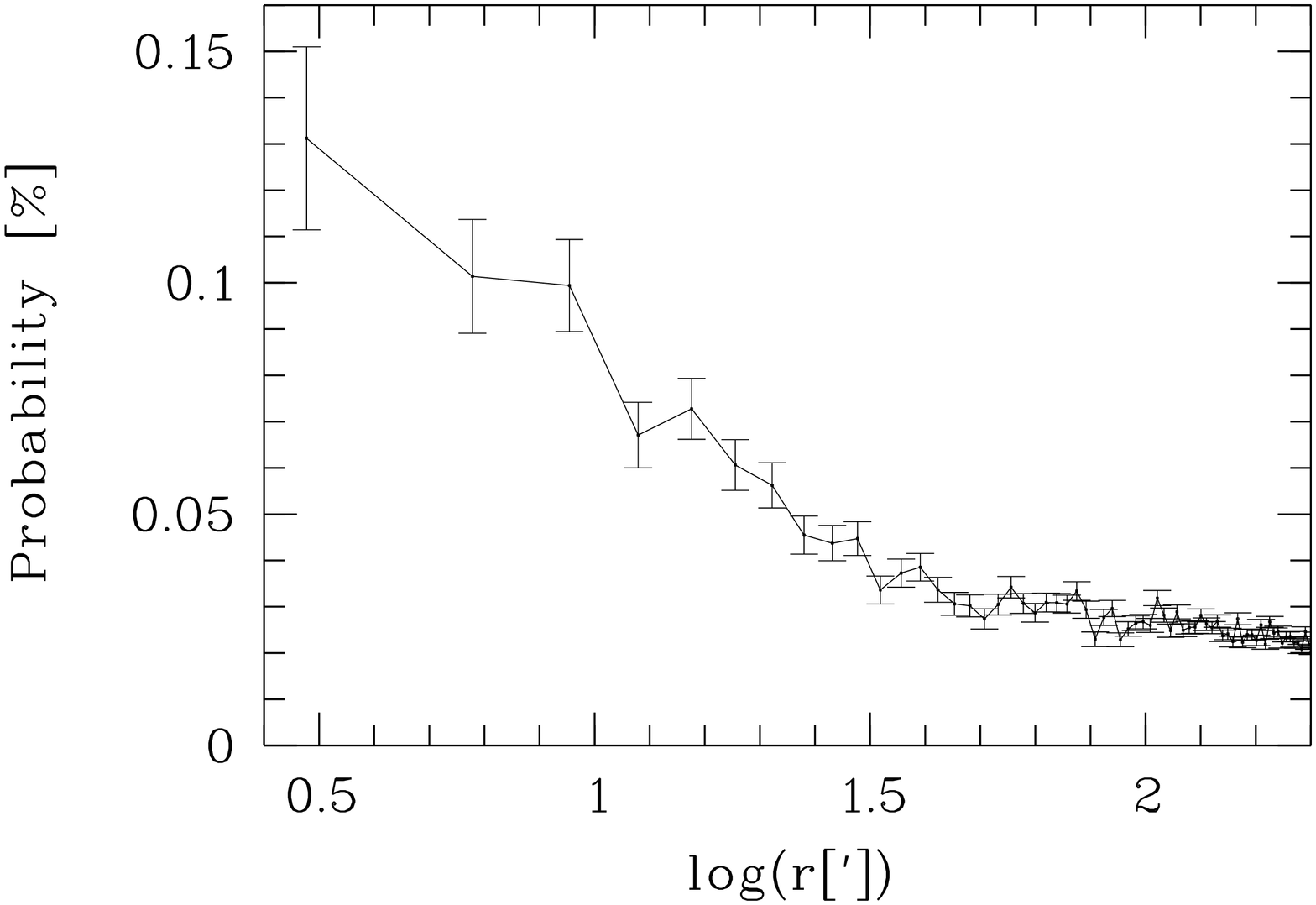}
\hfill
\includegraphics[height=8.5cm,angle=-90,bb=20 30 550 770]{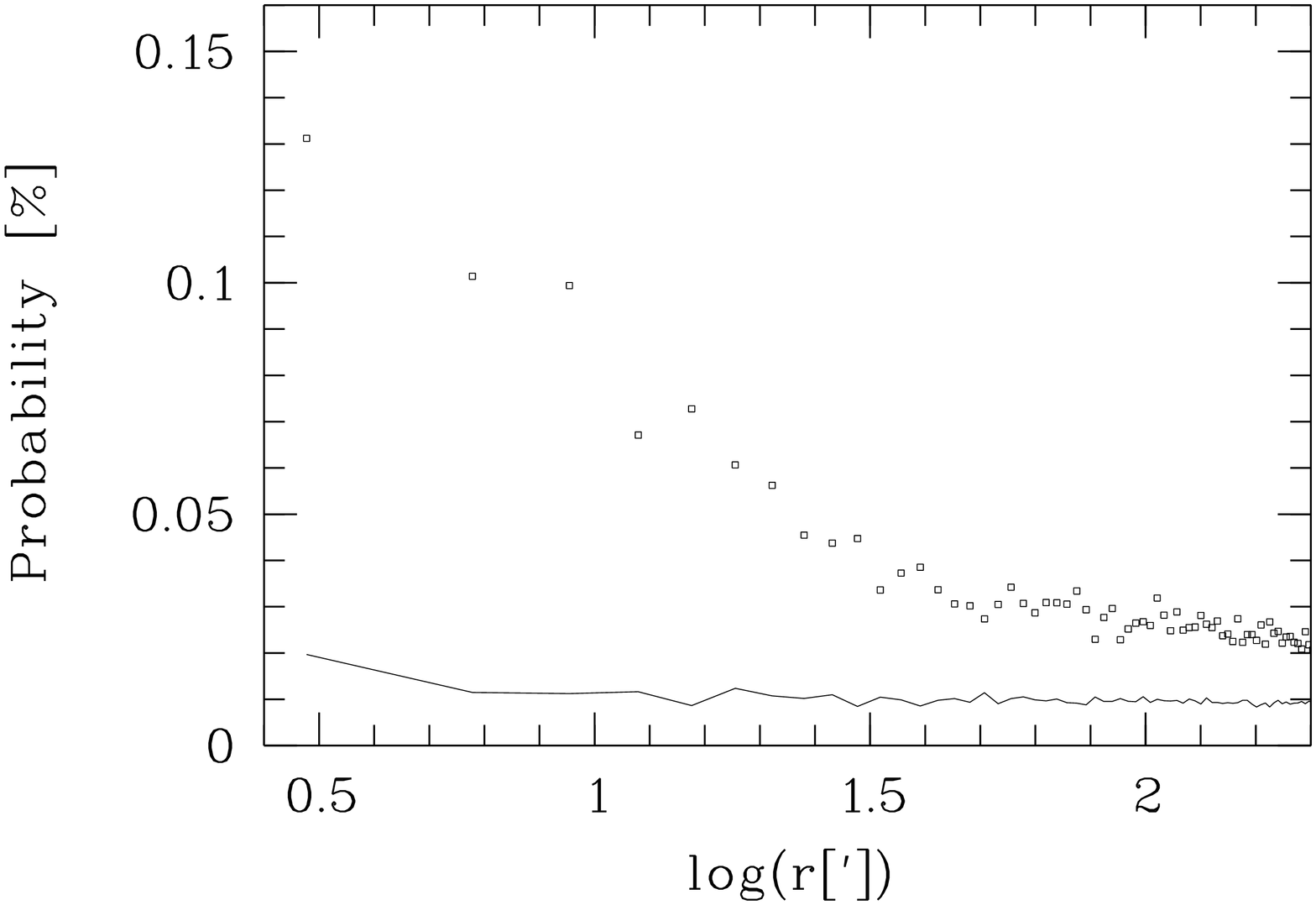}
\\
\includegraphics[height=8.5cm,angle=-90,bb=20 30 550 770]{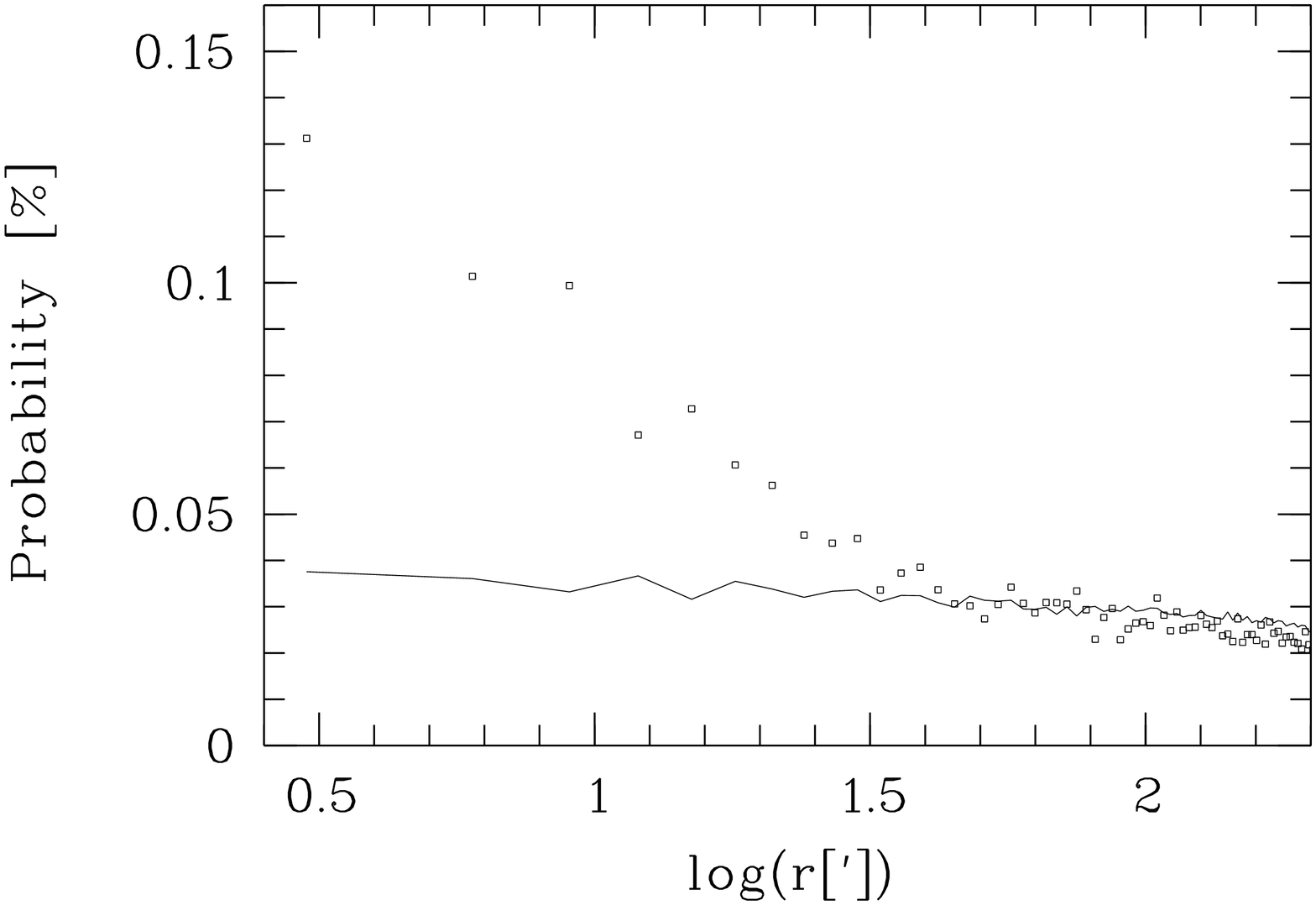}
\hfill
\includegraphics[height=8.5cm,angle=-90,bb=20 30 550 770]{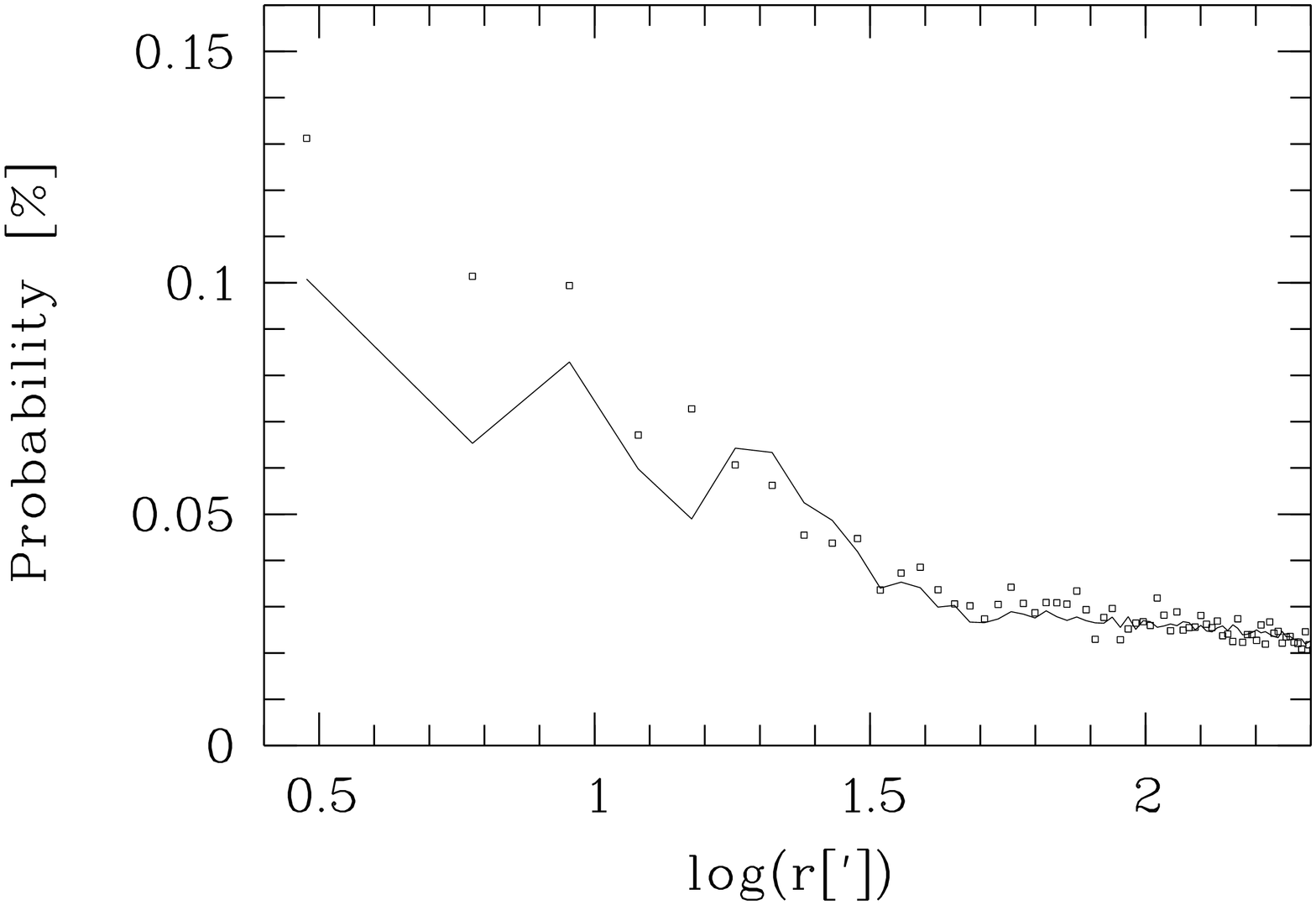}
\caption{\label{probability} Probability $P(r)$ to find pairs of clusters with a
separation $r$, normalised to the area and number of clusters. {\bf Upper left:}
Probability for the observational sample. Only star clusters which are more than
60$^\circ$ away from the Galactic Centre are included. Error bars are
statistical errors based on the number of clusters taken into account for each
individual datapoint. {\bf Upper right:} Probability distribution for a
homogeneous distribution of star clusters (solid line). Over plotted as dots are
the observational datapoints (shown without error bars). {\bf Lower left:}
Probability distribution for a model assuming a homogeneous distribution of 400
clusters. The remaining objects are homogeneously distributed along the Galactic
Plane and possess a Gaussian distribution with a width of 5$^\circ$
perpendicular to it. {\bf Lower right:} Probability distribution assuming a
homogeneous distribution of 400 clusters. 600 clusters are grouped in pairs of
two. These pairs, as well as the remaining clusters possess a homogeneous
distribution along the Galactic Plane and Gaussian distribution with a width of
5$^\circ$ perpendicular to the Galactic Plane. Within the pairs the clusters are
distributed homogeneously within 0.7$^\circ$.}
\end{figure*}

In the upper left panel of Fig.\,\ref{probability} we plot $P(r)$ for our
sample of clusters. There are two distinct regions in the diagram. For
separations larger than about 0.7$^\circ$ we find a linear decreasing trend of
$P$ with $\log{r}$. At smaller $r$ the probability rises much steeper with
decreasing radius. If we use only the 632 known OpCls to determine $P(r)$, in
order to investigate if the contamination of $\sim 50$\,\% influences the
distribution, we obtain the same qualitative behaviour but with a slightly
larger scatter due to the smaller number of clusters. Indeed all sub-samples of
clusters (manually detected objects, automatically detected objects, known
OpCls, new cluster candidates) show the same qualitative and quantitative
behaviour. This might be due to the fact that contaminating objects, i.e.
random density enhancements, are not expected to show any clustering, and thus
contribute only a constant to the plot shown in Fig.\,\ref{probability}. In any
case, the results of the following analysis are unaffected by contamination.

How does this observed distribution compare with models for the distribution of
the same number of clusters? The most simple model would be a homogeneous
distribution of the clusters in the investigated area. The resulting
probability plot for this model is shown in the upper right panel of
Fig.\,\ref{probability}. Note that we determined $P(r)$ as the average of ten
different homogeneous model distributions to minimise the scatter. The same
number of repeats is also used for all other models investigated in this
section. Obviously the homogeneous model is not appropriate, which is also
apparent in  Fig.\,\ref{spatialdistribution}. In contrast to the observational
sample, this model shows a probability that is almost independent of the
distance $r$, and also much smaller. This is due to the fact that a homogeneous
distribution leads on average to larger distances between the star clusters. In
fact an ideal homogeneous distribution would result in $\mathcal N_i(r)$ being
proportional to the separation $r$, and therefore a constant value for $P(r)$
(see Eq.\,\ref{prob_eq}). The good agreement of our homogeneous model
distribution with a constant value, reflects the high quality of the random
number generator used. The larger deviation for very small $r$ is due to a very
small number of cluster pairs at such a small separation. 

Can we obtain better agreement with a more realistic approach? From the
histograms in Fig.\,\ref{histolb} we can deduce the general properties of the
cluster distribution. There is a homogeneous component of about 400 clusters in
the investigated area. The remaining objects show a more or less homogeneous
distribution along the Galactic Plane (in the region more than 60$^\circ$ away
from the Galactic Centre), and a Gaussian distribution with a width of
5$^\circ$ perpendicular to the Galactic Plane. The lower left panel of
Fig.\,\ref{probability} shows the probability distribution for this model. For
$r$ larger than 0.7$^\circ$ the model matches the probability of the
observational data quite well. It reproduces the linear decline of $P$ with
$\log{r}$, and also the absolute values for $P$ are in agreement to the
observations. At small $r$ the model continues to show the same linear trend,
in contrast to the increase in the slope in the observational sample.

What causes the difference between the model and the observations for smaller
separations? Apparently at these shorter distances it is much more probable to
find another cluster than predicted by the simple model, obtained from the
histograms of the cluster distribution. This is clear evidence for strong,
local clustering of star clusters. We incorporate this in our models by again
using 400 clusters as homogeneous component. Additionally,  we sort 600 of the
remaining clusters in pairs of two. Those pairs have a size of 0.7$^\circ$, in
agreement with the change of slope in the probability distributions. Within the
0.7$^\circ$ box the coordinates of the two clusters are distributed
homogeneously. All pairs as well as all clusters that are not part of the
homogeneous component are distributed as in the above model (homogeneous along
and Gaussian, with a width of 5$^\circ$, perpendicular to the Galactic Plane).
We experimented with the number of clusters in each group, the group size, and
the number of clusters that are part of groups. The size of the group
determines the radius where we see a change in the slope of $P(r)$. If we put
more than two clusters in each group, the slope at small $r$ is much larger
than in the observations. The number of groups influences the absolute value of
$P(r)$.

This model is now in very good agreement with our observational sample. It
predicts very well the correct absolute values for the probability at all
separations. The change in the slope at 0.7$^\circ$ is also captured. At very
small $r$ (below 3'),  there is still some discrepancy between model and
observations, which might be due to sub structuring on very small scales. In
total, however, our simple model approach yields surprisingly good agreement
between simulated and observed cluster frequency. All simulations without
pairing (i.e. clustering) of objects are clearly not sufficient. This provides
strong evidence for clustering of star clusters on scales of the order of
0.7$^\circ$. Assuming a mean distance of 1-2\,kpc of the  clusters, we find
that the clustering of star clusters appears on scales of approximately
12-25\,pc. Since this is in the same order as the size of typical star forming
molecular  clouds, from which clusters are believed to form, this might
indicate that the observed distribution of clusters still contains information
about the birthplaces of stars in the Galaxy. 

\section{Classification of new cluster candidates}
\label{chapter4}

\subsection{Morphological analysis}
\label{propertyanalysis}

For every cluster candidate in our sample we fit the radial star density 
profile in the three filters JHKs at the completeness limit of 2MASS
(determined locally for every cluster). Note that we include all objects with a
quality flag better than 'E', corresponding to a photometric accuracy better
than three times the signal-to-noise ratio. We used the King-Profile (King
\cite{1962AJ.....67..471K}) 
\begin{equation}
\rho (r) = \rho_b + \rho_c \cdot r_{cor}^2 \cdot \left[ \left( r_{cor}^2 + r^2
\right)^{-\frac{1}{2}} - \left( r_{cor}^2 + r_{tid}^2 \right)^{-\frac{1}{2}}
\right]^2
\label{kingprofile}
\end{equation}
for the star density $\rho$ to determine the core and tidal radius ($r_{cor}$,
$r_{tid}$) and the central and background star density ($\rho_c$, $\rho_b$).

Integrating the cluster star density $\rho (r) - \rho_b$ from zero to the tidal
radius and substituting $x \equiv r_{tid} / r_{cor}$, we can determine the total
number $N$ of stars in the cluster by:
\begin{equation}
N = \pi \cdot \rho_c \cdot r_{cor}^2 \cdot \left[ \ln(1+x^2) - 4 + \frac{4
\cdot \sqrt{1+x^2} + x^2}{1+x^2} \right].
\label{eq_stcl}
\end{equation}

This total number of stars depends on the magnitude down to which stars are
included in the star density map, i.e. the local completeness limit near the
cluster $m_{cl}$. To compare the number of cluster members between different
star clusters, we have to convert $N$ to a common magnitude limit $m_{all}$.
This number $N_c$ will be used in Sect.\,\ref{classification} to distinguish
between OpCls and GlCls, and can in principle be estimated for each cluster
using: 
\begin{equation}
N_c = N \cdot 10^{-\frac{C}{2.5} \cdot (m_{cl} - m_{all})}.
\label{totnumber}
\end{equation}
The variable $C$ in Eq.\,\ref{totnumber} denotes an unknown parameter. It
describes how the number of stars in a cluster changes with stellar luminosity,
i.e. a combination of the mass and luminosity function. It depends e.g. on the
distance, type, age and extinction of the cluster, as well as on $m_{cl}$. The
variable $C$ only scales Eq.\,\ref{totnumber}; thus the particular value of it
is not important and we will set it to one for the analysis in Sect.
\ref{classification}. Note that different values only marginally influence the
discrimination of the two cluster types. 

\subsection{Cluster parameters and uncertainties}
\label{qualflagdet} 

We list the determined cluster parameters for the new cluster candidates in
Tables\,\ref{sourcelist} and \ref{sourcelist_sde} in the Appendix. The tables
will be available online only and contain the following columns: 1) Unique ID
for each new cluster candidate; 2) Galactic Longitude 3) Galactic Latitude; 4)
Right Ascension (J2000); 5) Declination (J2000); 6) core radius in the H-band;
7) tidal radius in the H-band; 8) central star density in the H-band; 9)
Intensity contrast -- central star density/sqrt(background star density) in the
H-band; 10) Number of stars in the cluster in the H-band; 11) Corrected number
of stars in the cluster in the H-band (see Sect.\,\ref{propertyanalysis}); 12)
log $R_N$ (see Sect.\,\ref{classification} for details). 13) Quality flag (see
below). 14) Name of possible known cluster with erroneous coordinates in
SIMBAD.

What are the uncertainties of the parameters? The cluster coordinates are
determined by a Gaussian fit of the star density peak. The star density maps
are created by counting stars in 3.5\,arcmin sized boxes with an oversampling
of ten. Considering this, the accuracy of the positions is in the order of the
pixel size of our maps, i.e. 20" or 0.005\,degrees. However, in cases of
extended or not centrally condensed clusters, the uncertainty could reach
0.01\,degrees. The error of the core and tidal radii can be estimated as
follows: we determine the scatter of the fit radii from the mean value in all
three filters JHK. As error for the radii we assume a value that is larger than
the scatter of the best 2/3 of the objects. For the core radius we find that
2/3 of the objects posses a scatter of less than 15\,\%. In case of the tidal
radius the situation is worse. The best 2/3 of the objects posses a scatter of
less than 33\,\%. Hence for many objects the tidal radius is not well
determined. 

To give a measure to judge the quality of the radial star density profile fit
and the reliability of the cluster parameters of each new cluster candidate we
introduce a quality flag. This flag consists of two parts: a) an integer number
ranging from zero to six, indicating how many of our quality tests are negative
for this particular object; b) a six digit binary number, allowing to identify
which of the quality tests are negative. The following bits are used: 1) the
cluster is detected automatically (0) or manually (1); 2) the scatter of the
three fitted core radii is larger than 30\,\% (1) or smaller (0); 3) the
scatter of the three fitted tidal radii is larger than 66\,\% (1) or smaller
(0); 4) the contrast of the central to background star density
(Icentral/sqr(Iback)) in all filters is larger than five (0) or not (1); 5) the
ratio of tidal to core radius in all three filters is inbetween 3 and 45 (0) or
not (1); 6) the core radius in all three filters is larger than twice the error
of the coordinates, i.e. 0.01\,degrees (0) or not (1).

\subsection{Classification: OpCl vs. GlCl}
\label{classification}

\begin{figure}
\includegraphics[height=8.2cm,angle=-90,bb=35 35 540 760]{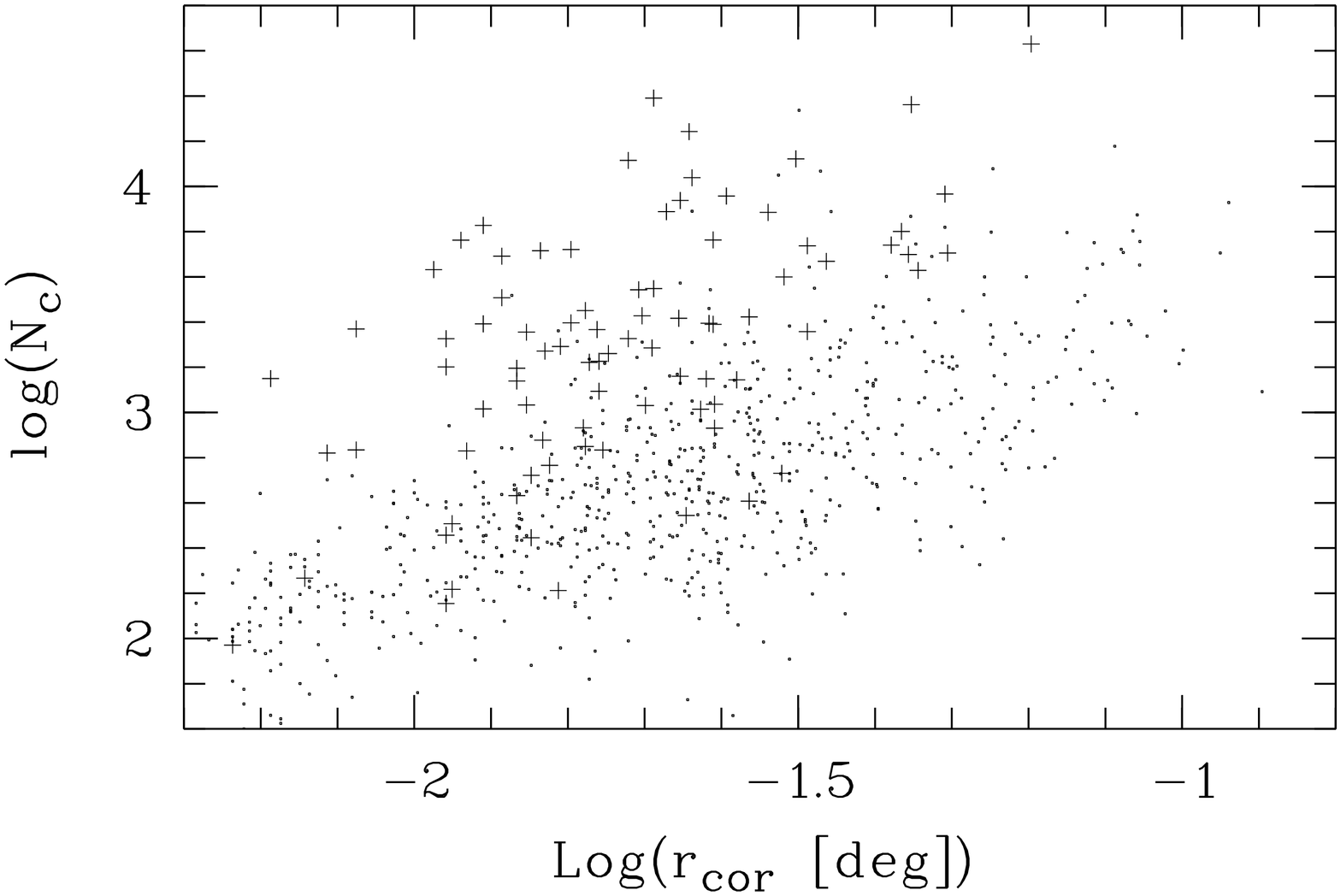}
\\
\includegraphics[height=8.2cm,angle=-90,bb=35 35 540 760]{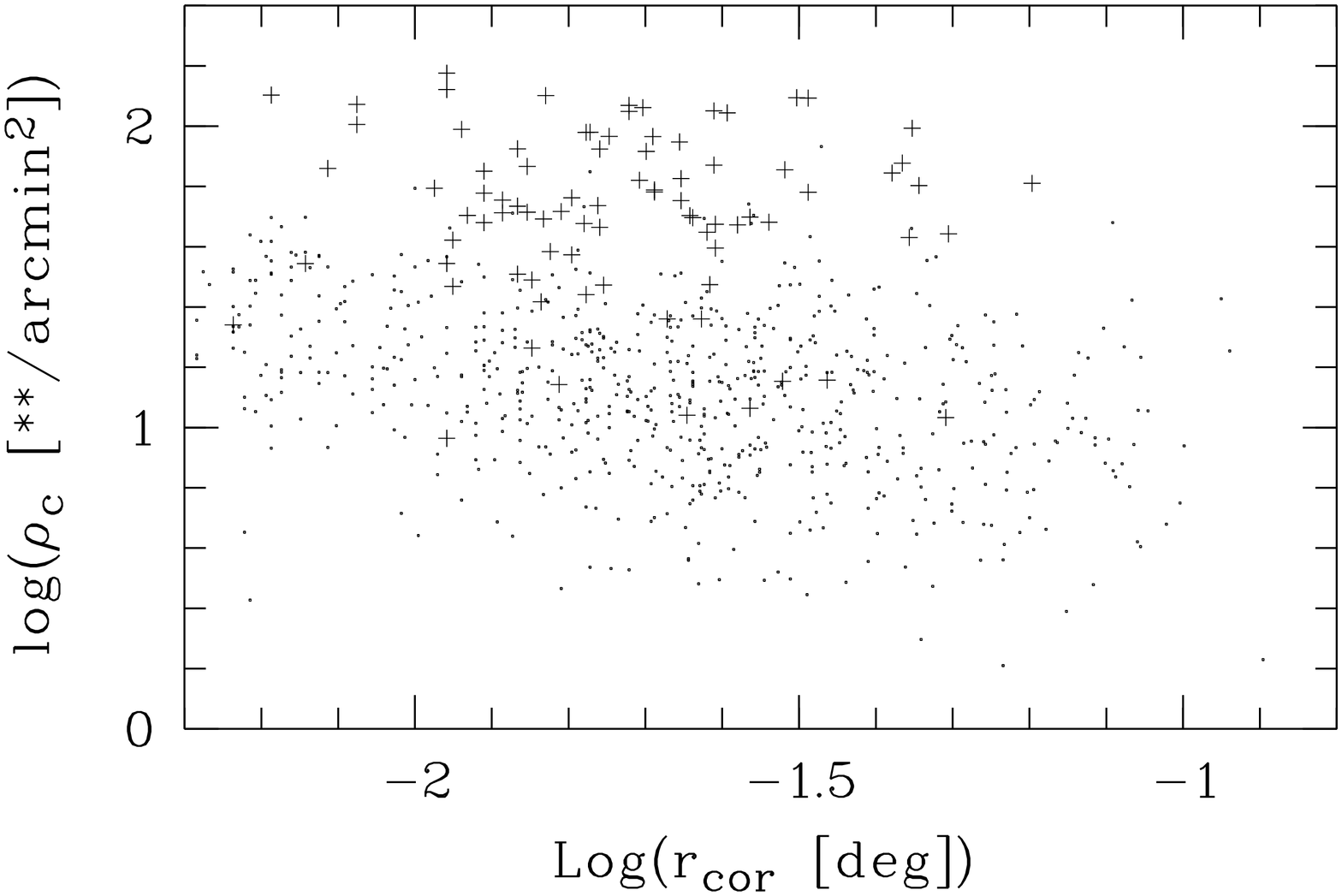}
\\
\includegraphics[height=8.2cm,angle=-90,bb=35 35 540 760]{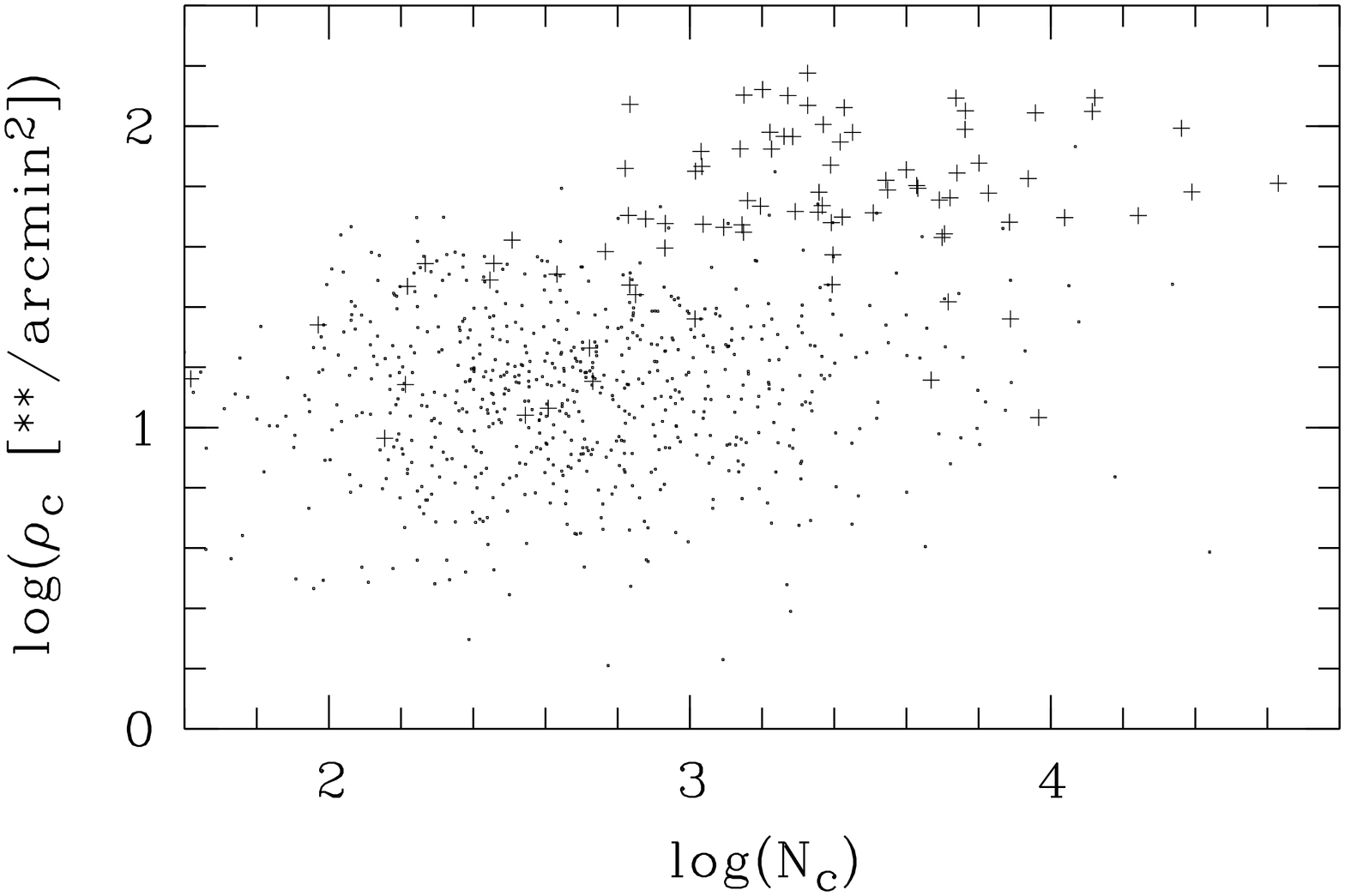}
\caption{\label{properties} Example plots of cluster properties (H-band data)
for know OpCls (dots) and known GlCls (plus signs) in our object sample, which
were used to classify the new cluster candidates. {\bf Top:} Corrected number of
stars vs. the core radius. {\bf Middle:} Central star density vs. the core
radius. {\bf Bottom: } Central star density vs. the corrected number of stars.
It clearly can be seen that the OpCls occupy a different part of the parameter
space than most of the GlCls.}    
\end{figure}

Our large and homogeneous sample of known star clusters gives us the
opportunity to distinguish between OpCls and GlCls among the new cluster
candidates based on  statistical differences in their morphological properties,
as determined in Sect. \ref{propertyanalysis}. In particular we are interested
in identifying the most probable new GlCl candidates within the vast number of
unknown objects. For this purpose we investigated all possible combinations of
two morphological  properties (determined as described in
Sect.\,\ref{propertyanalysis}) for the known clusters in our sample. We then
selected the combinations that showed the best discrimination of OpCls and
GlCls. The three best combinations for this purpose are plotted in
Fig.\,\ref{properties} (presented are data from H-band images), which shows how
the corrected number of stars  $N_c$, the core radius and the central star
density relate to each other for the known OpCls  (dots) and GlCls (plus signs)
in our sample. It is evident in Fig.\,\ref{properties}  that the two types of
clusters occupy different parts of the parameter space, even  if some GlCls are
well within the OpCl regime.

The three parameter combinations shown in Fig.\,\ref{properties} are available
in three different filters JHKs, resulting in nine different parameter
combinations  that we can use to distinguish OpCls and GlCls. The high
redundancy in all these different relationships as well as the large samples of
known clusters in our  database ensures a reliable analysis of the statistical
differences between OpCls and GlCls.

In each plot we count how many known GlCls and OpCls are in a box around each
star cluster candidate in our sample. The box has a size of 0.3 in logarithmic
units. This number is then normalised by the total number of known clusters in
our sample. These ratios are a measure for the probability that this cluster
candidate is a GlCl ($P^{Gl}_{plot_i}$) or OpCl ($P^{Op}_{plot_i}$). Combining
these individual probabilities from all plots using 
\begin{equation} 
P^{Op/Gl} = \left( \prod\limits_{i=1}^9 P^{Op/Gl}_{plot_i} \right)^{1/9} 
\end{equation} 
we determined a measure for the probability of each new cluster
candidate to be a GlCl ($P^{Gl}$) or OpCl ($P^{Op}$). 

Almost all new cluster candidates are most likely OpCls. We note, however, that
the sample of known OpCls is very incomplete, especially towards distant
clusters and objects with few stars. Compared to this, the sample of GlCls is
almost complete, only about 10\,\% are believed to be missing towards the
Galactic Centre (Ivanov et al. \cite{2005A&A...442..195I}). Therefore, the
probability $P^{Op}$, is a lower limit. The probability $P^{Gl}$, however,
opens the possibility to identify the best candidates for the last missing
GlCls in our Galaxy. Given the incompleteness of the OpCl sample, however,
$P^{Gl}$ has to be considered an upper limit.

\subsection{Best new GlCl candidates}
\label{bestglcldet}

\begin{figure}
\includegraphics[height=8.5cm,angle=-90,bb=20 20 550 780]{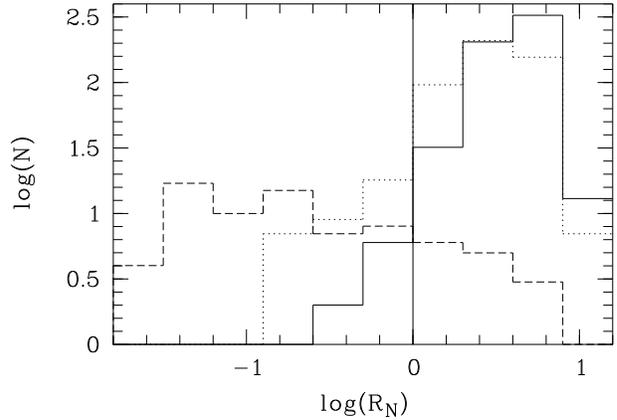}
\caption{\label{probratio} Distribution of the measure $R_N$ for our sample of
clusters. The value of $R_N$ gives an indication of what type the star cluster
is. We plot logarithmic histograms vs. the logarithm of the measure $R_N$. The
dividing line at $\log{R_N} = 0$ separates the area of GlCls (negative
$\log{R_N}$) and OpCls (positive $\log{R_N}$). The solid histogram represents
the new cluster candidates in our sample, the dashed histogram the known GlCls
and the dotted histogram the known OpCls.}    
\end{figure}

Here we investigate the feasibility of our method to determine a measure for
the probability that new cluster candidates in our sample are GlCls or OpCls
and to select new globular cluster candidates. For each cluster, we determined
the ratio $P^{Op} / P^{Gl}$. If we had no information at all about the
morphological properties, this ratio would be equal to the ratio of the number
of the two types of known clusters in our sample. Hence, by defining $R_N
\equiv (P^{Op} / P^{Gl}) / (681/86)$ we obtain an easy measure for each cluster
to be GlCl or OpCl. If $\log{R_N}$ is negative, the object is probably a GlCl,
otherwise it is an OpCl.

In Fig.\,\ref{probratio} we plot the histogram for our cluster candidates vs. 
$\log{R_N}$, our measure for the classification of clusters. The solid
histogram represents the new cluster candidates in our sample, the dashed
histogram the known GlCls and the dotted histogram the known OpCls. The
vertical solid line marks the (statistical) borderline between globular and
open clusters. Note that in Fig.\,\ref{probratio} objects which have a
$P^{Gl}$-value of zero (and hence undetermined $R_N$) are missing. Those
objects, however, are  most probably OpCls. As can be seen in the figure, most
known clusters are classified correctly using this approach: Out of the 86
known GlCls, only 14 (16\,\%) have $\log{R_N}>0$ and thus fail our criterium.
For OpCl, the agreement is even better. Only 36 out of 681 (5\,\%) OpCls are
falsely classified by our approach.

There are nine new cluster candidates in our sample that possess a value of
$R_N$ less than one, and are hence promising GlCl candidates (see Table
\ref{bestnewglcl}). Given the ratio of known GlCls and OpCls in  this parameter
range (72/36) and a over-all contamination rate of about 50\,\% of the 1021 new
cluster candidates (see Sect. \ref{overall}) we estimate that about 25\,\%,
i.e. 2-3 of those candidates are likely to be new GlCls. However, this estimate
is subject to the uncertainties in the completeness of the known OpCls in our
sample and the contamination of our new candidates. We also note that all new
GlCl candidates have rather low (albeit negative) values of $\log{R_N}$, close
to the OpCl regime, and all but one have a quality flag of two or higher. A
visual inspection of 2MASS JHK colour images shows no clear indication of a
star cluster for all but the cluster candidate 1716 (see Fig.\,\ref{twonew2}).
As discussed in Sect. \ref{reliab}, the impression of colour images can be
misleading, because new clusters might be dominated by faint stars, which do
not strongly affect the visual (subjective) appearance of the cluster.

Nevertheless, it will be justified to systematically investigate the new
cluster candidates with the smallest $\log{R_N}$-values e.g. with deep and
high-resolution NIR imaging. This is particularly interesting in areas close to
the Galactic Plane and the Galactic Centre, were extinction hampers the
detection of GlCls at optical wavelength. The recently discovered GlCl
Glimpse\,C01 for example, which has $\log{R_N}$\,=\,-0.851 according to our
analysis, is situated in this area ($l$\,=\,31.31, $b$\,=\,-0.10).

\begin{table}
\caption{\label{bestnewglcl} Positions of the nine new cluster candidates in our
sample that are classified as most probably being a GlCl. The
$\log{R_N}$-values for all other new candidates and their properties can
be found in the Appendix in Table\,\ref{sourcelist}.}
\begin{center}
\begin{tabular}{crrccc}
Cluster & $l\,[^\circ]$ & $b\,[^\circ]$ & $\log{R_N}$ & \multicolumn{2}{c}{Qual.
Flag} \\ \hline 
0001 & 0.029   &  3.474 & -0.43 & 2 & 100010 \\
0002 & 0.048   &  3.440 & -0.62 & 2 & 100010 \\
0055 & 17.992  & -0.278 & -0.03 & 4 & 110110 \\
0089 & 29.491  & -0.981 & -0.15 & 2 & 100010 \\
1603 & 298.222 & -0.507 & -0.11 & 0 & 000000 \\
1703 & 325.788 &  0.124 & -0.32 & 3 & 100110 \\
1716 & 329.792 & -1.589 & -0.11 & 2 & 100010 \\
1755 & 348.246 &  0.482 & -0.19 & 2 & 100010 \\
1767 & 352.602 & -2.168 & -0.05 & 2 & 100001 \\
\end{tabular}
\end{center}
\end{table}

\section{Conclusions}

We have used star density maps determined from the 2MASS point source catalogue
to obtain a complete sample of star clusters in the entire Galactic Plane ($|b|
< 20^\circ$). We used a combination of automated searches for local density
enhancements and manual detection to obtain our cluster sample. In total 1788
cluster candidates are detected by our method. There are 86 previously known
GlCls and 681 OpCls among our source sample. The remainder of 1021 objects are
new star cluster candidates.

An analysis of the spatial distribution of the clusters shows that the majority
of the new candidates are similarly distributed as the previously known OpCls.
There is, however, a component of roughly 500 objects, that appears to be
homogeneously distributed. These sources are therefore most likely local star
density enhancements and not real star clusters. We estimate the contamination
rate of our candidate sample to be about 50\,\%.

We determined how the probability of finding star cluster pairs changes with
separation. Using a simple model for the two dimensional cluster distribution,
we were able to reproduce the observations. The model assumes a homogeneously
distributed component of clusters (i.e. the contamination of our sample), a
component that is homogeneously distributed along and Gaussian distributed
perpendicular to the Galactic Plane, and a fraction of clusters in close-by
pairs. These close-by pairs of star clusters are unconditionally required to
reproduce the observational star cluster distribution providing strong evidence
for clustering of star clusters on small scales. The size of the groups of
0.7$^\circ$ indicates a locally enhanced cluster density on scales of
12-25\,pc, a typical size of molecular clouds.

The large and homogeneous sample of known OpCls and GlCls allowed us to use a
statistical approach to classify the new cluster candidates. This was done by
means of cluster properties obtained by fitting the radial star density profile
of each object by a King-Profile. The measure for classification obtained by us
classifies only 16\,\% of the known GlCls and only 5\,\% of the known OpCls
wrongly. According to our criterium, there are nine promising GlCl candidates
among our new cluster candidates, from which 2-3 are likely to be real GlCl. A
systematic detailed investigation of the most promising GlCl candidates is
suggested in order to identify possible new Galactic Globular Clusters in our
sample.

\section*{acknowledgements}

We gratefully acknowledge stimulating discussions with Helmut Meusinger about
parts of this paper. D.\,Froebrich received funding by the Cosmo Grid project,
funded by the Program for Research in Third Level Institutions under the
National Development Plan and with assistance from the European Regional
Development Fund.  This publication makes use of data products from the Two
Micron All Sky Survey, which is a joint project of the University of
Massachusetts and the Infrared Processing and Analysis Center/California
Institute of Technology, funded by the National Aeronautics and Space
Administration and the National Science Foundation. This research has made use
of the SIMBAD database, operated at CDS, Strasbourg, France.

\label{lastpage}

\begin{appendix}

\onecolumn
\section{Properties of probable cluster candidates}

\renewcommand{\tabcolsep}{2pt}
\begin{longtable}{crrrrrrccrrrrrp{3cm}}
\caption{\label{sourcelist} Properties of the probable clusters in our sample
of newly detected star cluster candidates. The columns contain the following:
(1) Unique identifier for each cluster candidate; (2) Galactic Longitude in
degrees (typical error 0.01\degr); (3) Galactic Latitude in degrees (typical
error 0.01\degr); (4) Right Ascention (J2000) (typical error 0.01\degr); (5)
Declination (J2000) (typical error 0.01\degr); (6) Core radius in degrees as
fitted to the H-band radial star density profile (typical error 15\,\%); (7)
Tidal radius in degrees as fitted to the H-band radial star density profile
(typical error 33\,\%); (8) Central star density in stars per square arcminutes
as fitted to the H-band radial star density profile; (9) Star density contrast
in the H-band, determined as ratio of the central star density and the square
root of the background star density. Values of -99 indicate objects with a
fitted background star density of zero; (10) Number of stars in the cluster as
obtained using Eq.\,\ref{eq_stcl} and the star density profile in the H-band;
(11) Corrected number of stars in the cluster determined using
Eq.\,\ref{totnumber} at an apparent magnitude of 15 in the H-band; (12) The
measure $\log{R_N}$ as determined in Sect.\,\ref{bestglcldet}. Values of 99
indicate cluster candidates with $P^{Gl} = 0.$; (13) Quality Flag determined
from our analysis of the radial density profile fit in
Sect.\,\ref{qualflagdet}; (14) Name of possible known cluster with
erroneous coordinates in SIMBAD.}  \\ 
\noalign{\smallskip}
\hline
\noalign{\smallskip} 
\multicolumn{1}{c}{(1)} &
\multicolumn{1}{c}{(2)} &
\multicolumn{1}{c}{(3)} &
\multicolumn{1}{c}{(4)} &
\multicolumn{1}{c}{(5)} &
\multicolumn{1}{c}{(6)} &
\multicolumn{1}{c}{(7)} &
\multicolumn{1}{c}{(8)} &
\multicolumn{1}{c}{(9)} &
\multicolumn{1}{c}{(10)} &
\multicolumn{1}{c}{(11)} &
\multicolumn{1}{c}{(12)} &
\multicolumn{2}{c}{(13)} &
\multicolumn{1}{c}{(14)} \\
\noalign{\smallskip} 
\hline
\noalign{\smallskip}
ID & \multicolumn{1}{c}{$l$} & \multicolumn{1}{c}{$b$} &
\multicolumn{2}{c}{RA/DEC (J2000)} & $r_{\rm core}^h$ &  
$r_{\rm tid}^h$ & $I_{\rm cent}^h$ & Con.$^h$ & 
$N^h$ & $N_c^h$ & $\log{R_N}$ & \multicolumn{2}{c}{Qual. Flag} & Poss. Known\\ 
\noalign{\smallskip}
 & \multicolumn{1}{c}{[\degr]} & \multicolumn{1}{c}{[\degr]} &
 \multicolumn{1}{c}{[h:m:s]} & \multicolumn{1}{c}{[\degr:':'']} &
 \multicolumn{1}{c}{[\degr]} & \multicolumn{1}{c}{[\degr]} & [**/'$^2$] & & & &
 & & & Cluster \\   
\noalign{\smallskip}
\endfirsthead
\multicolumn{15}{c}{\tablename\ \thetable{} -- continued from previous
page} \\ 
\noalign{\smallskip}
\hline
\noalign{\smallskip}
ID & \multicolumn{1}{c}{$l$} & \multicolumn{1}{c}{$b$} &
\multicolumn{2}{c}{RA/DEC (J2000)} & $r_{\rm core}^h$ &  
$r_{\rm tid}^h$ & $I_{\rm cent}^h$ & Con.$^h$ & 
$N^h$ & $N_c^h$ & $\log{R_N}$ & \multicolumn{2}{c}{Qual. Flag} & Poss. Known \\ 
\noalign{\smallskip}
 & \multicolumn{1}{c}{[\degr]} & \multicolumn{1}{c}{[\degr]} &
 \multicolumn{1}{c}{[h:m:s]} & \multicolumn{1}{c}{[\degr:':'']} &
 \multicolumn{1}{c}{[\degr]} & \multicolumn{1}{c}{[\degr]} & [**/'$^2$] & & & &
 & & & Cluster \\   
\noalign{\smallskip}
\hline
\noalign{\smallskip}
\endhead
\noalign{\smallskip}
\hline
\noalign{\smallskip}
\multicolumn{15}{r}{Continued on next page} \\
\noalign{\smallskip}
\endfoot
\hline
\endlastfoot
\hline
\noalign{\smallskip}
0001 & 0.029   & 3.474 		& 17:32:22 & -27:03:39 & 0.026 & 0.872 & 17 & 5  & 1654 & 7221 & -0.379 & 2 & 100010 &  \\
0002 & 0.047   & 3.44 		& 17:32:32 & -27:03:51 & 0.089 & 0.178 & 32 & 10 & 6534 & 28520& -0.55  & 2 & 100010 &  \\
0022 & 6.182   & 0.843 		& 17:56:28 & -23:11:34 & 0.039 & 0.078 & 16 & 4  & 642  & 4439 & 0.142  & 3 & 100110 &  \\
0023 & 6.584   & 0.782 		& 17:57:35 & -22:52:32 & 0.006 & 0.119 & 25 & 7  & 119  & 686  & 0.28   & 3 & 110001 &  \\
0031 & 8.906   & -0.268 	& 18:06:29 & -21:22:33 & 0.08  & 3.349 & 15 & 5  & 15152& 41732& 99.0   & 2 & 000110 &  \\
0035 & 9.687   & 0.762          & 18:04:16 & -20:11:27 & 0.016 & 0.048 & 11 & 3  & 107  & 512  & 99.0   & 4 & 110110 &  \\
0039 & 10.25   & 0.32 		& 18:07:05 & -19:54:57 & 0.014 & 0.072 & 18 & 5  & 219  & 795  & 0.401  & 3 & 100101 &  \\
0046 & 13.351  & 0.066 		& 18:14:20 & -17:19:19 & 0.053 & 0.106 & 16 & 5  & 1142 & 2868 & 0.42   & 3 & 100110 &  \\
\end{longtable}

\clearpage
\newpage

\section{Properties of possible cluster candidates}

\renewcommand{\tabcolsep}{2pt}
\begin{longtable}{crrrrrrccrrrrrp{3cm}}
\caption{\label{sourcelist_sde} Properties of the possible clusters in our
sample of newly detected star cluster candidates. The columns contain the
following: (1) Unique identifier for each cluster candidate; (2) Galactic
Longitude in degrees (typical error 0.01\degr); (3) Galactic Latitude in
degrees (typical error 0.01\degr); (4) Right Ascention (J2000) (typical error
0.01\degr); (5) Declination (J2000) (typical error 0.01\degr); (6) Core radius
in degrees as fitted to the H-band radial star density profile (typical error
15\,\%); (7) Tidal radius in degrees as fitted to the H-band radial star
density profile (typical error 33\,\%); (8) Central star density in stars per
square arcminutes as fitted to the H-band radial star density profile; (9) Star
density contrast in the H-band, determined as ratio of the central star density
and the square root of the background star density. Values of -99 indicate
objects with a fitted background star density of zero; (10) Number of stars in
the cluster as obtained using Eq.\,\ref{eq_stcl} and the star density profile
in the H-band; (11) Corrected number of stars in the cluster determined using
Eq.\,\ref{totnumber} at an apparent magnitude of 15 in the H-band; (12) The
measure $\log{R_N}$ as determined in Sect.\,\ref{bestglcldet}. Values of 99
indicate cluster candidates with $P^{Gl} = 0.$; (13) Quality Flag determined
from our analysis of the radial density profile fit in
Sect.\,\ref{qualflagdet}; (14) Name of possible known cluster with
erroneous coordinates in SIMBAD.}  \\ 
\noalign{\smallskip}
\hline
\noalign{\smallskip} 
\multicolumn{1}{c}{(1)} &
\multicolumn{1}{c}{(2)} &
\multicolumn{1}{c}{(3)} &
\multicolumn{1}{c}{(4)} &
\multicolumn{1}{c}{(5)} &
\multicolumn{1}{c}{(6)} &
\multicolumn{1}{c}{(7)} &
\multicolumn{1}{c}{(8)} &
\multicolumn{1}{c}{(9)} &
\multicolumn{1}{c}{(10)} &
\multicolumn{1}{c}{(11)} &
\multicolumn{1}{c}{(12)} &
\multicolumn{2}{c}{(13)} &
\multicolumn{1}{c}{(14)} \\
\noalign{\smallskip} 
\hline
\noalign{\smallskip}
ID & \multicolumn{1}{c}{$l$} & \multicolumn{1}{c}{$b$} &
\multicolumn{2}{c}{RA/DEC (J2000)} & $r_{\rm core}^h$ &  
$r_{\rm tid}^h$ & $I_{\rm cent}^h$ & Con.$^h$ & 
$N^h$ & $N_c^h$ & $\log{R_N}$ & \multicolumn{2}{c}{Qual. Flag} & Poss. Known \\ 
\noalign{\smallskip}
 & \multicolumn{1}{c}{[\degr]} & \multicolumn{1}{c}{[\degr]} &
 \multicolumn{1}{c}{[h:m:s]} & \multicolumn{1}{c}{[\degr:':'']} &
 \multicolumn{1}{c}{[\degr]} & \multicolumn{1}{c}{[\degr]} & [**/'$^2$] & & & &
 & & & Cluster \\   
\noalign{\smallskip}
\endfirsthead
\multicolumn{15}{c}{\tablename\ \thetable{} -- continued from previous
page} \\ 
\noalign{\smallskip}
\hline
\noalign{\smallskip}
ID & \multicolumn{1}{c}{$l$} & \multicolumn{1}{c}{$b$} &
\multicolumn{2}{c}{RA/DEC (J2000)} & $r_{\rm core}^h$ &  
$r_{\rm tid}^h$ & $I_{\rm cent}^h$ & Con.$^h$ & 
$N^h$ & $N_c^h$ & $\log{R_N}$ & \multicolumn{2}{c}{Qual. Flag} & Poss. Known \\ 
\noalign{\smallskip}
 & \multicolumn{1}{c}{[\degr]} & \multicolumn{1}{c}{[\degr]} &
 \multicolumn{1}{c}{[h:m:s]} & \multicolumn{1}{c}{[\degr:':'']} &
 \multicolumn{1}{c}{[\degr]} & \multicolumn{1}{c}{[\degr]} & [**/'$^2$] & & & &
 & & & Cluster \\   
\noalign{\smallskip}
\hline
\noalign{\smallskip}
\endhead
\noalign{\smallskip}
\hline
\noalign{\smallskip}
\multicolumn{15}{r}{Continued on next page} \\
\noalign{\smallskip}
\endfoot
\hline
\endlastfoot
\hline
\noalign{\smallskip}
0009 & 1.864   & -9.522         & 18:28:29 & -31:53:47 & 0.005 & 0.25  & 16 & 6  & 69   & 91   & 0.802  & 4 & 100111 &  \\
0010 & 2.149   & 19.617         & 16:40:49 & -16:01:09 & 0.023 & 0.227 & 7  & 7  & 330  & 250  & 0.845  & 0 & 000000 &  \\
0014 & 4.406   & 16.753         & 16:55:42 & -15:59:40 & 0.009 & 0.25  & 10 & 8  & 122  & 92   & 0.779  & 3 & 110001 &  \\
0018 & 5.338   & 5.407          & 17:37:41 & -21:33:28 & 0.019 & 0.096 & 13 & 4  & 276  & 1586 & 99.0   & 3 & 110100 &  \\
0019 & 5.52    & 6.08           & 17:35:39 & -21:02:49 & 0.026 & 0.103 & 10 & 4  & 315  & 2180 & 99.0   & 2 & 000110 &  \\
0025 & 7.535   & 5.648          & 17:41:43 & -19:34:15 & 0.018 & 0.884 & 9  & 4  & 510  & 3529 & 99.0   & 3 & 100110 &  \\
0027 & 7.783   & 8.451          & 17:32:12 & -17:53:04 & 0.018 & 0.194 & 13 & 5  & 381  & 459  & 0.666  & 2 & 100100 &  \\
0033 & 9.359   & 10.139         & 17:29:41 & -15:40:11 & 0.086 & 0.172 & 9  & 4  & 1710 & 1560 & 99.0   & 4 & 011110 &  \\
\end{longtable}

\end{appendix}

\end{document}